\documentclass{aa}
\usepackage{graphicx}
\usepackage{txfonts}
\usepackage{aalongtable}
\usepackage{lscape}

\def\intgr{{\it INTEGRAL}}

\begin{document}

\title{XMM-Newton and INTEGRAL analysis of the Ophiuchus cluster of galaxies}

\author{J. Nevalainen \inst{1}
          \and
        D. Eckert\inst{2}
           \and
        J. Kaastra\inst{3}
           \and
        M. Bonamente\inst{4}
           \and
        K. Kettula\inst{1}
          }

   \offprints{J. Nevalainen}

   \institute{Observatory, University of Helsinki, Finland\\
              \email{Jukka.H.Nevalainen@helsinki.fi}
          \and
      ISDC Data Centre for Astrophysics, Geneva Observatory, University of Geneva, Switzerland\\ 
          \and
          SRON Netherlands Institute for Space Research, Netherlands\\
          \and
       University of Alabama in Huntsville, Huntsville, USA ; NASA National Space and Technology Center, Huntsville, USA
     }    

   \date{Received; accepted}

\abstract
{}
{We investigated the non-thermal hard X-ray emission in the Ophiuchus cluster of galaxies. Our aim is to characterise
the physical properties of the non-thermal component and its interaction with the cosmic microwave background.}
{We performed spatially resolved spectroscopy and imaging using XMM-Newton data to model the
thermal emission. Combining this with INTEGRAL ISGRI data, we modelled the 0.6--140 keV band total emission in the central
7 arcmin region.}
{The models that best describe both PN and ISGRI data contain a power-law component with a photon index in a 
range 2.2--2.5.  This component produces $\sim$10\% of the total flux in the 1--10 keV band. The pressure of the 
non-thermal electrons is $\sim$1\% of that of the thermal electrons.
Our results support the scenario whereby a relativistic electron population, which produces the recently detected radio mini-halo in Ophiuchus, also produces the hard X-rays via inverse compton scattering of the CMB photons.
The best-fit models imply a differential momentum spectrum of the relativistic electrons with a slope of 3.4--4.0 and
a magnetic field strength B=0.05--0.15 $\mu$G. The lack of evidence for a recent major merger in the 
Ophiuchus centre allows the possibility that the relativistic electrons are produced by turbulence or hadronic 
collisions.}
{}


\authorrunning{J. Nevalainen et al.}
\titlerunning{XMM and INTEGRAL analysis of Ophiuchus}

\maketitle

\section{Introduction}
Non-thermal hard X-ray emission has been detected in several clusters of galaxies over the past few years
(see Rephaeli et al., 2008 for a recent review). Since the detections have remained at the level of a few $\sigma$,
many models still remain as valid explanations for the phenomenon. The most popular explanation is the 
inverse compton scattering of the cosmic microwave background photons with the relativistic electrons in the cluster
(e.g. Sarazin et al., 1988).
In this model, the CMB photon ends up in the hard X-ray band since its energy increases by a factor of $\sim 10^8$ via the 
scattering.  
In primary models, the relativistic electrons are originally thermal electrons that have been accelerated by 
a cluster merger (e.g. Sarazin \& Lieu, 1998) or turbulence (e.g. Brunetti et al., 2001; Brunetti et al., 2004).
In secondary models, the acceleration comes from hadronic collisions (e.g. Dennison, 1980; Pfrommer \& Ensslin, 2004).

Recently a radio mini-halo was detected in the centre of the Ophiuchus cluster with the VLA at 1.4 GHz (Govoni et al., 2009; 
Murgia et al., 2009). This proved the existence of a population of relativistic electrons in Ophiuchus.
Ophiuchus is a hot (T$\sim$9 keV) nearby (z=0.028) cluster located close to the Galactic centre 
(l $\sim 1^{\circ}$ , b $\sim  9^{\circ}$).
Consistently, INTEGRAL detected excess emission over the thermal component at a 4--6$\sigma$ level
in the 20--80 keV energy band in the Ophiuchus cluster (Eckert et al., 2008). 
The analysis was lacking a sensitive instrument for modelling 
the thermal component at lower energies, but the excess was  nevertheless consistent with having a non-thermal origin. 

Our aim in this work is to improve the modelling of the emission in the Ophiuchus centre utilising spatially resolved
XMM-Newton spectroscopy. In this work we use the EPIC instruments PN for spectral analysis and MOS2 for imaging of 
the central 7 arcmin region of the Ophiuchus cluster.
Because of the limitations of the spatial capability of INTEGRAL, we cannot exclude regions of complex temperature structure.
Rather, we examine the spatially resolved XMM-Newton data in order to obtain an accurate model for the thermal emission
to be used later with the INTEGRAL data. We also update the INTEGRAL analysis with additional data.

\begin{figure*}[tb]
\includegraphics[width=18cm,angle=0]{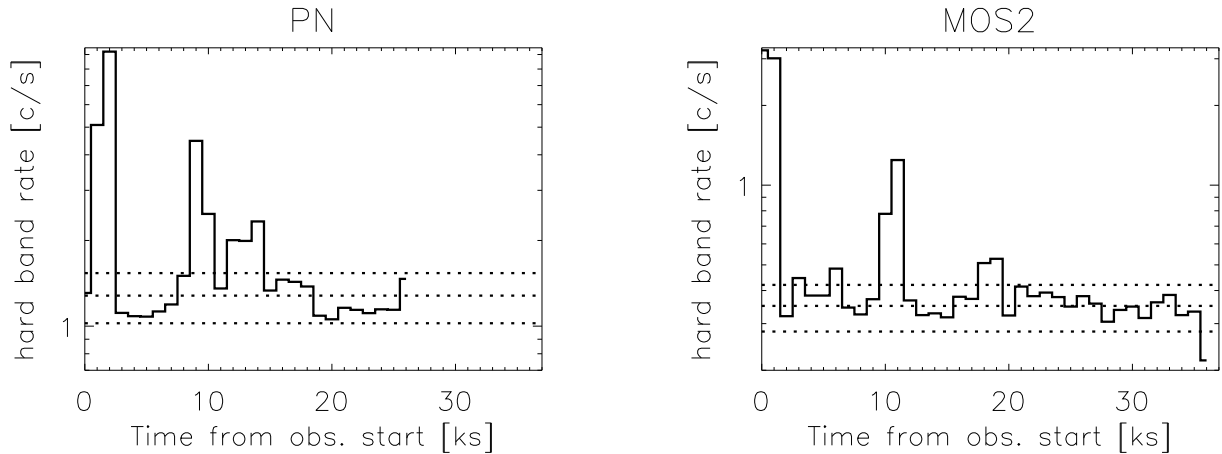}
\vspace{-6cm}
\caption{The histogram shows the light curve of Ophiuchus in the full FOV in 1000 ks time bins in the hard band (E $>$ 10 
keV for PN, left panel and  E $>$ 9.5 keV for MOS2, right panel). The dotted horizontal lines show 
the adopted quiescent level. 
\label{lc.fig}}
\end{figure*}

\section{XMM-Newton data reduction}
We observed the Ophiuchus cluster with XMM-Newton on September 2, 2007, during revolution 1416. 
The observation ID is 0505150101.
The PN instrument was operated in extended full frame mode and the EPIC instruments used the medium 
optical filter. 

\subsection{Processing the data}
We processed the raw data with the SAS version xmmsas\_20080701\_1801-8.0.0 tools epchain and emchain with the
default parameters in order to produce the event files. We used the latest calibration information in October 2008.
We also generated the simulated out-of-time event file, which we 
later used to subtract the events registered during readout of a CCD from PN spectra. We filtered the event files 
excluding bad pixels and CCD gaps. We further filtered the event files including only patterns 0--4 (PN) and 0--12 (MOS).
We used the evselect-3.58.7 tool to extract spectra, images, and light curves, while excluding the regions contaminated 
by bright point sources. We used the rmfgen-1.55.1 and arfgen-1.75.5
tools to produce energy redistribution files and the auxiliary response files.

\subsection{Background}
\label{back}
We extracted the hard band ( E$>$10 keV for PN,  E$>$9.5 keV for MOS2) light curves of the full field-of-view region
in order to examine the behaviour of the particle background (see Fig. \ref{lc.fig}).
We found that while the fraction of the exposure time containing the flares was small, 
the quiescent level exceeded that of the closed filter sample (Nevalainen et al., 2005) by a factor of $\sim$2.
Thus, instead of using the quiescent level from blank sky compilations we defined it as a level exceeding the light 
curve minimum by  20\%. We accepted such periods, during which the hard band count rate was within $\pm$20\% of the 
quiescent level. This results in 16 ks and 28 ks of useful exposure time for PN and MOS2, respectively.

\begin{figure}
\includegraphics[width=9cm,angle=0]{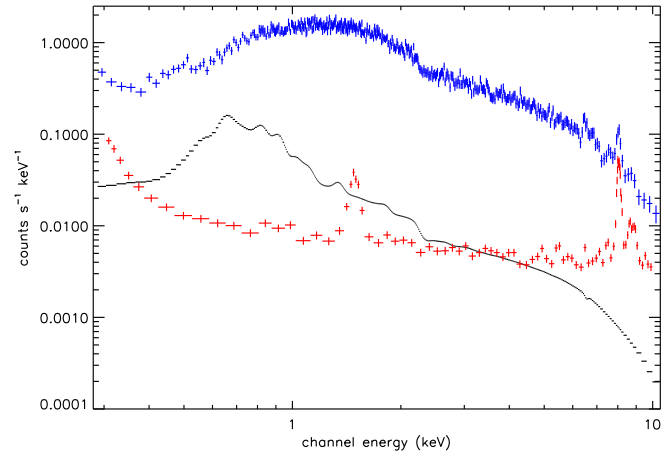}
\caption{The total PN emission spectrum obtained within the 6.5--7.0 arcmin annulus, containing the cluster and 
background components, is shown with blue crosses. 
Also shown separately are the background components, i.e.
the detector background (red crosses) and the sky background (black curve). 
\label{bkg.fig}}
\end{figure}

Since Ophiuchus is very bright and its spectrum is relatively hard, there is some cluster signal even at a photon energy 
of 10 keV (see Fig. \ref{bkg.fig}). At higher energies (E$>$14 keV for PN; E$>$ 10.5 keV for MOS2), the spectrum of Ophiuchus is consistent with 
that of the scaled closed filter sample, by a factor of 1.77 and 1.93 for PN and MOS2, respectively. 
Thus, in the following we use the closed filter spectrum, multiplied by the above factor to remove the particle 
background from the data.        

Since Ophiuchus fills the whole FOV we cannot obtain a local estimate for the sky background. 
Neither are the blank sky compilations useful, since the Galactic emission and absorption are very high due to  
Ophiuchus being located close to the Galactic Centre. We thus used the HEASARC tool based on ROSAT All Sky Survey data
to obtain the spectrum of the Galaxy. We used fields with centre within 0.5--1.0 degree around the Ophiuchus centre. 
We modelled the spectrum using an unabsorbed MEKAL for the Local Hot Bubble, and an absorbed MEKAL + power-law
to account for the emission of the Galactic halo and the cosmic X-ray background. We kept 
the photon index fixed to 1.41 as found in the measurements by De Luca et al. (2004). 
The best-fit MEKAL temperatures were in the range 0.1--0.2 keV.
The obtained value and 1$\sigma$ statistical uncertainty interval for NH is 0.46 [0.39--0.66] $\times 10^{21}$ cm$^{-2}$,
much higher than that derived with the radio measurements of HI at 21 cm (Kalberla et al. 2005). 
We will address this issue in detail in Section \ref{spectralfitting}.

Compared to the background, the cluster emission is strong even at the outermost annulus (with radii 6.5--7.0 arcmin)
used in this work, where the cluster is the faintest (see Fig. \ref{bkg.fig}). 
In order to keep the detector and sky background emission below 10\% of the cluster signal, we limited our spectral 
analysis to photons with energy in the range 0.6--7.4 keV in the following.

\section{INTEGRAL data reduction}
An analysis of INTEGRAL data on the cluster was already presented in Eckert et al. (2008). We aim here at complementing 
the results presented in this paper by the addition of new data and a very careful study of the systematic effects which
might affect the results.

\subsection{Data selection and spectral extraction}
\begin{figure}
\includegraphics[width=7cm,angle=-90]{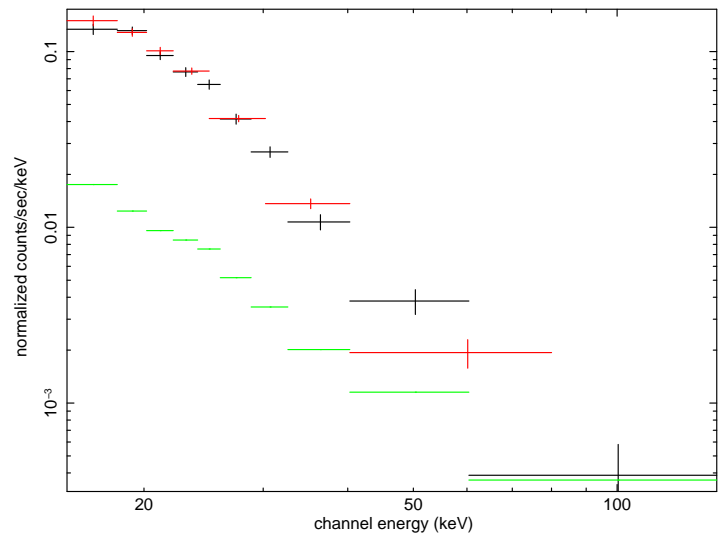}
\caption{\intgr/ISGRI spectrum of the Ophiuchus cluster extracted with 2 different methods (black: standard OSA spectral extraction tool, red: extraction from mosaic images). The green lines show the 2-$\sigma$ noise level estimated from 10 random positions within a radius of 2 degrees around the source.}
\label{noise}
\end{figure}

The IBIS/ISGRI instrument (Lebrun et al., 2003)  on board \intgr\  (Winkler et al., 2003) is a wide-field 
($29^\circ\times29^\circ$) coded-mask instrument operating in the 15-400 keV band. It is made of an array of 
$128\times128$ CdTe pixels and it features an angular resolution of 12 arcmin FWHM. Thanks to the observing strategy of 
\intgr, which spends a significant fraction of its observing time in the Galactic Bulge region, and to the large 
field-of-view (FOV), the available \intgr\ exposure time on the Ophiuchus cluster ($l=0.5^\circ,b=0.5^\circ$) is large. 
For this analysis, we used all publicly-available data from the \intgr\ archive, restricting to the observations where 
the 
source was at most $8^\circ$ offset from the pointing position, and we filtered out the data where the background rate 
was found to be more than 3 times larger than the average (e.g. because of solar flares). Overall, our data set comprises 
$\sim$1900 individual pointings, for a total of 3 Msec of high-quality data on the source. This allows us to detect the 
source with high significance ($53\sigma$) in the 15-140 keV band (see Fig. \ref{noise}).

To analyse the data, we used the standard Offline Scientific Analysis (OSA) package, version 7.0, distributed by ISDC 
(Courvoisier et al., 2003). The analysis of the ISGRI data in this region is made difficult by the presence of many point
sources in the field (e.g. the bright Low-Mass X-ray Binary Scorpius X-1), whose shadow pattern overlaps with that of the
Ophiuchus cluster. In order to take the contribution of other sources in the FOV into account, we prepared a catalogue 
comprising the 37 brightest sources detected in the mosaic images of the region, and used this catalogue as input to the 
spectral extraction tool, which then performs a multi-parameter fit of the shadow pattern of each sources to the raw 
detector images. In addition, it is known that the presence of screws and glue strips attaching the IBIS mask to the 
supporting structure can introduce systematic effects in the presence of very bright sources. To get rid of this 
systematic effect, we identified the mask areas where screws and glue are absorbing incoming photons, and we ignored the 
pixels which were illuminated by these areas for the three brightest sources in the field (Scorpius X-1, 4U 1700-377 and 
GRS 1758-258). As a cross check, we also extracted the ISGRI spectra directly by fitting the mosaic images, although this
method is known to be less reliable for flux estimates.
Except for the first bin (15-18 keV), where the ISGRI calibration is known to be uncertain due to the pixel 
low threshold, there is excellent agreement between the spectra extracted with the 2 methods.

\begin{figure*}[t]
\hbox{
\includegraphics[width=8cm,angle=0]{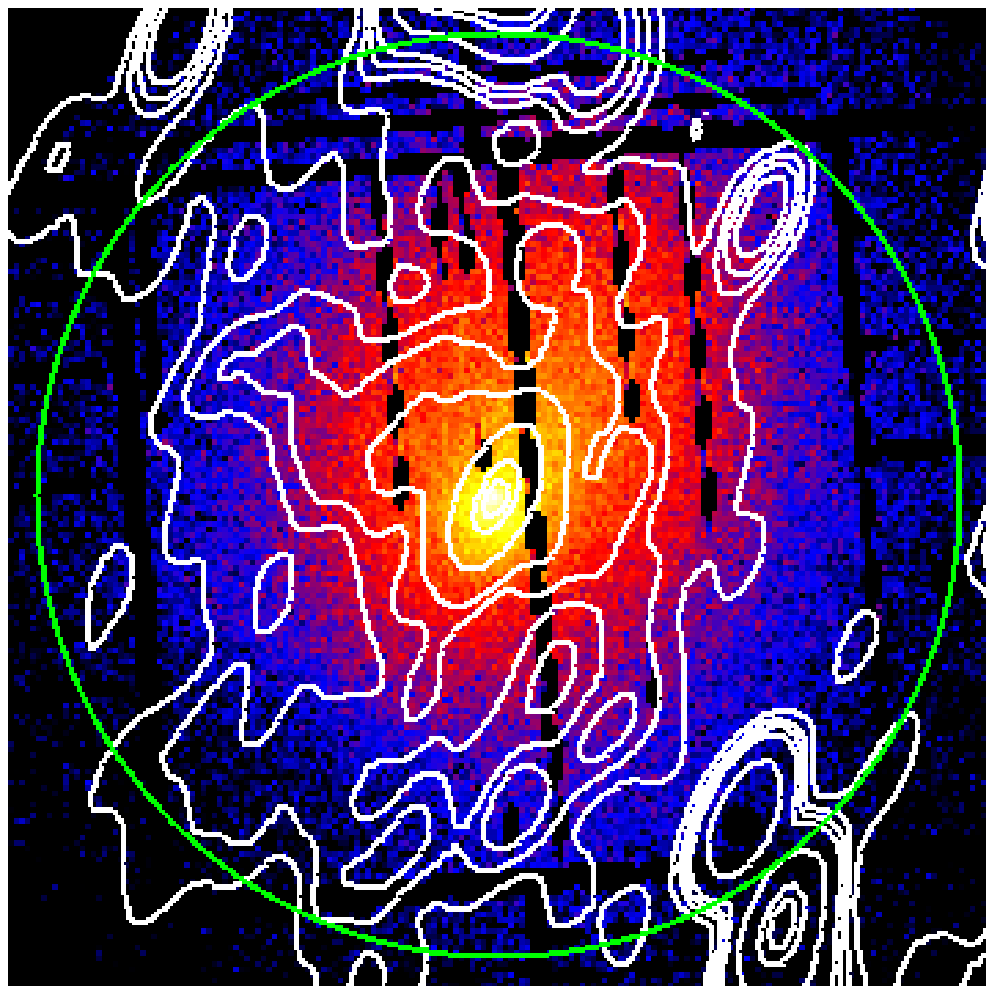}
\includegraphics[width=7.78cm,angle=0]{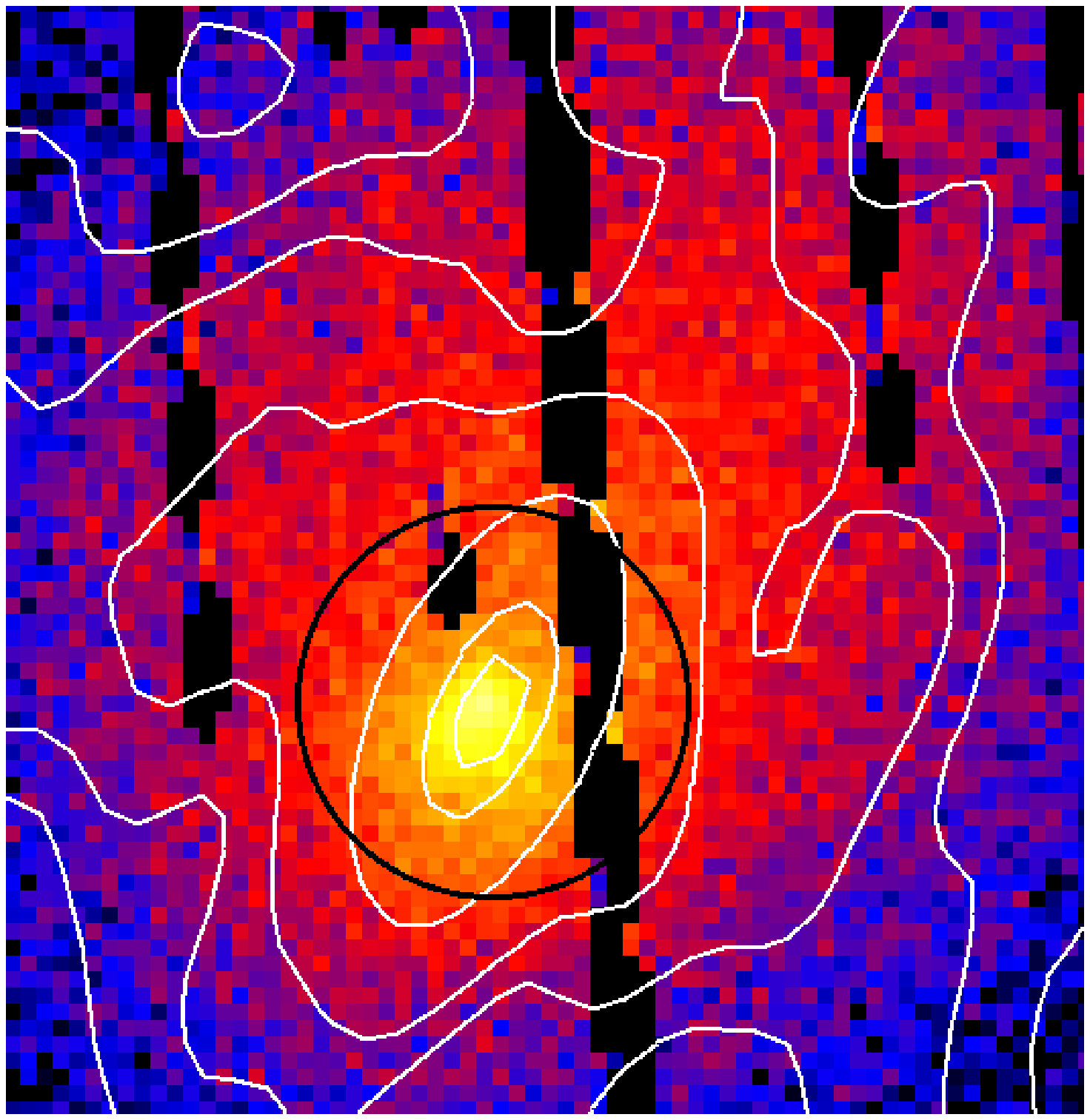}}
\caption{Vignetting-corrected MOS2 image of the Ophiuchus cluster centre in the 0.5--7.0 keV band. The pixel size is 5''.
The white curves show radio contours at 1.4 GHz (from Govoni et al., 2009).
In the left panel the green circle shows the extraction region (r=7 arcmin) for the XMM-Newton spectrum.
The right panel shows a close-up of the central region. The circle denotes the central r=1 arcmin (= 33 kpc) region.
\label{mos2image.fig}}
\end{figure*}

In order to estimate the contamination of the spectra by the bright objects in the FOV, we extracted ISGRI spectra at 10 
random positions within a radius of 2 degrees from the Ophiuchus cluster. As expected, no detection was obtained 
for any of the 10 observations and the data were distributed normally in all energy bins (see Fig. \ref{noise}). 
According to the latest 
\intgr\ cross-calibration report based on OSA 7.0 (Jourdain et al., 2008), the systematic uncertainties on the ISGRI RMF 
and ARF calibration are at 1\% level. As a conservative estimate for the systematic uncertainties on the energy 
calibration of ISGRI, a flat 2\% systematic error has been added to the spectra presented in Fig. \ref{noise}. In any 
case, we note that the level of statistical uncertainties is always significantly higher than 2\%, so the uncertainties 
on the energy calibration do not affect significantly the results.

For the detailed analysis of the spectra presented hereafter, we restrict to the spectrum extracted using the standard 
spectral extraction method, and we ignore the first energy bin (15-18 keV) because of uncertain calibration.

\section{XMM-Newton imaging analysis}
\label{spatres}
We used MOS2 for the imaging, since that EPIC instrument has the smallest fraction (4\%) of the central 7 arcmin region obscured by the dead 
area due to e.g. CCD gaps and bad pixels. We accumulated an image in the 0.5--7.0 keV band in 5 arcsec bins. 
We subtracted the particle background component from the imaging data using the closed cover compilation of Nevalainen et 
al. (2005). 
We then produced the exposure map using the SAS {\it eexpmap-4.4.1} tool and by dividing the 
imaging data with the exposure map, we produced a vignetting 
corrected surface brightness map. We further used the exposure map to produce a mask file for filtering out the data 
close to the CCD gaps. 

\subsection{Brightness distribution}
\label{surfbr}
The resulting image (Fig. \ref{mos2image.fig}) shows that the brightness distribution in the central region 
is elliptical, with an axis ratio of $\sim$0.7.
The location of the brightness peak is offset from the centroid of the elliptical brightness distribution by 
$\sim$1.5 arcmin, i.e. 50 kpc in the plane of the sky.
The X-ray morphology and the centroid shift do vary with time in the cluster merger simulations (e.g. Poole et al.,
2006) and thus this shift may be taken as an indication of merger activity in the Ophiuchus centre.
However, the variation due to merger details prohibits us to use the above values to estimate the time passed since the 
possible major merger in the centre of Ophiuchus.

There are some deviations from the azimuthally symmetric brightness distribution.
At 3.5--7 arcmin distance from the X-ray peak towards the south-east the surface brightness is lower than in the 
rest of the cluster at same radii. 
The edge-like feature of the brightness distribution at $\sim$ 1.2 arcmin (= 40 kpc) distance from the X-ray peak 
towards south and west coincides with the density discontinuity found in Chandra data (Markevitch et al., 2007), who 
interpreted the structure as a cold front. A detailed analysis of density discontinuities at small angular scales is
not warranted with XMM-Newton due to its relatively large PSF. 
The XMM-Newton data indicate that a surface brightness jump bigger 
than 10\% in the central 2 arcmin region towards south is excluded.

\begin{figure*}[t]
\hbox{
\includegraphics[width=8cm,angle=0]{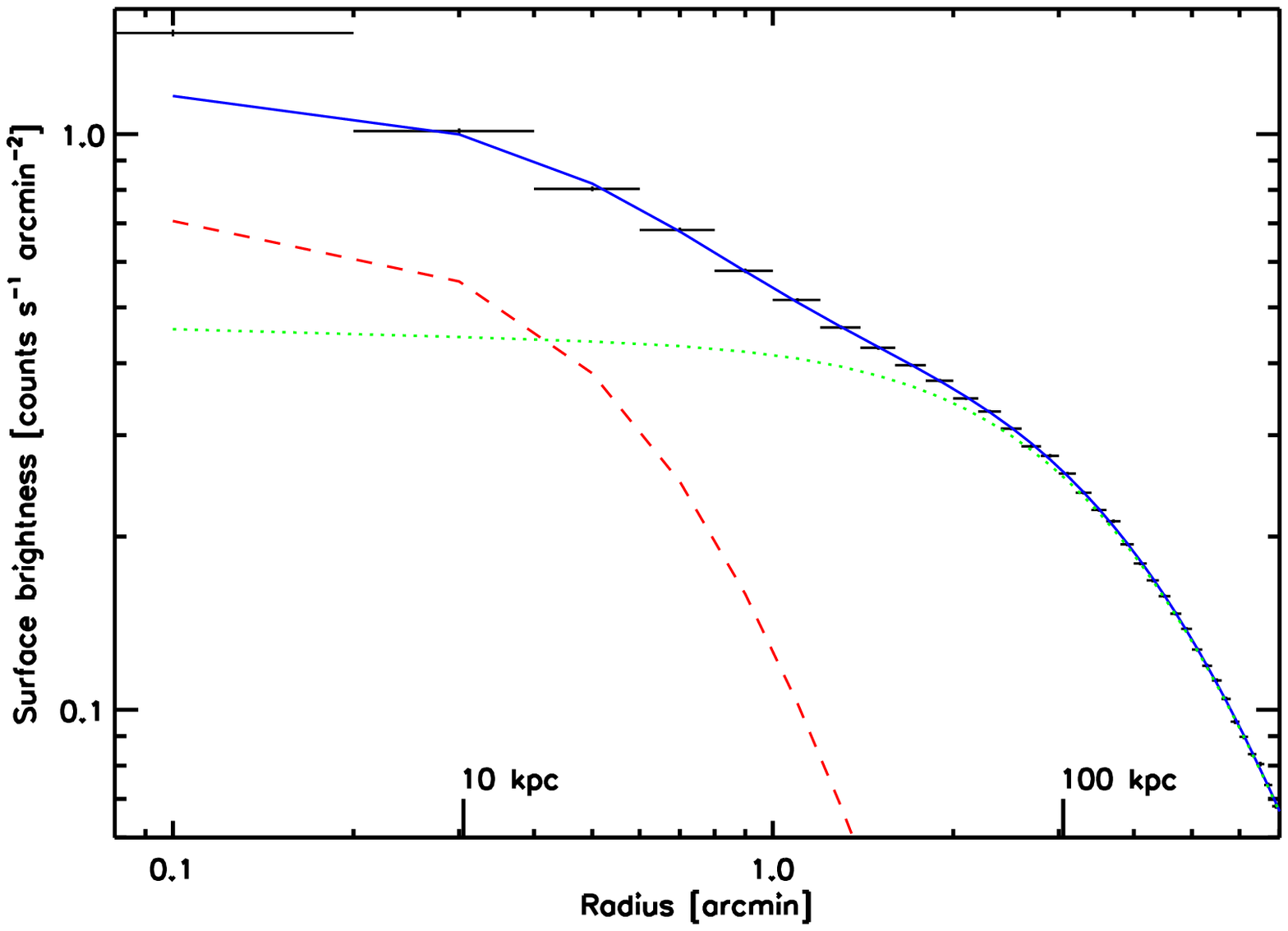}
\includegraphics[width=8cm,angle=0]{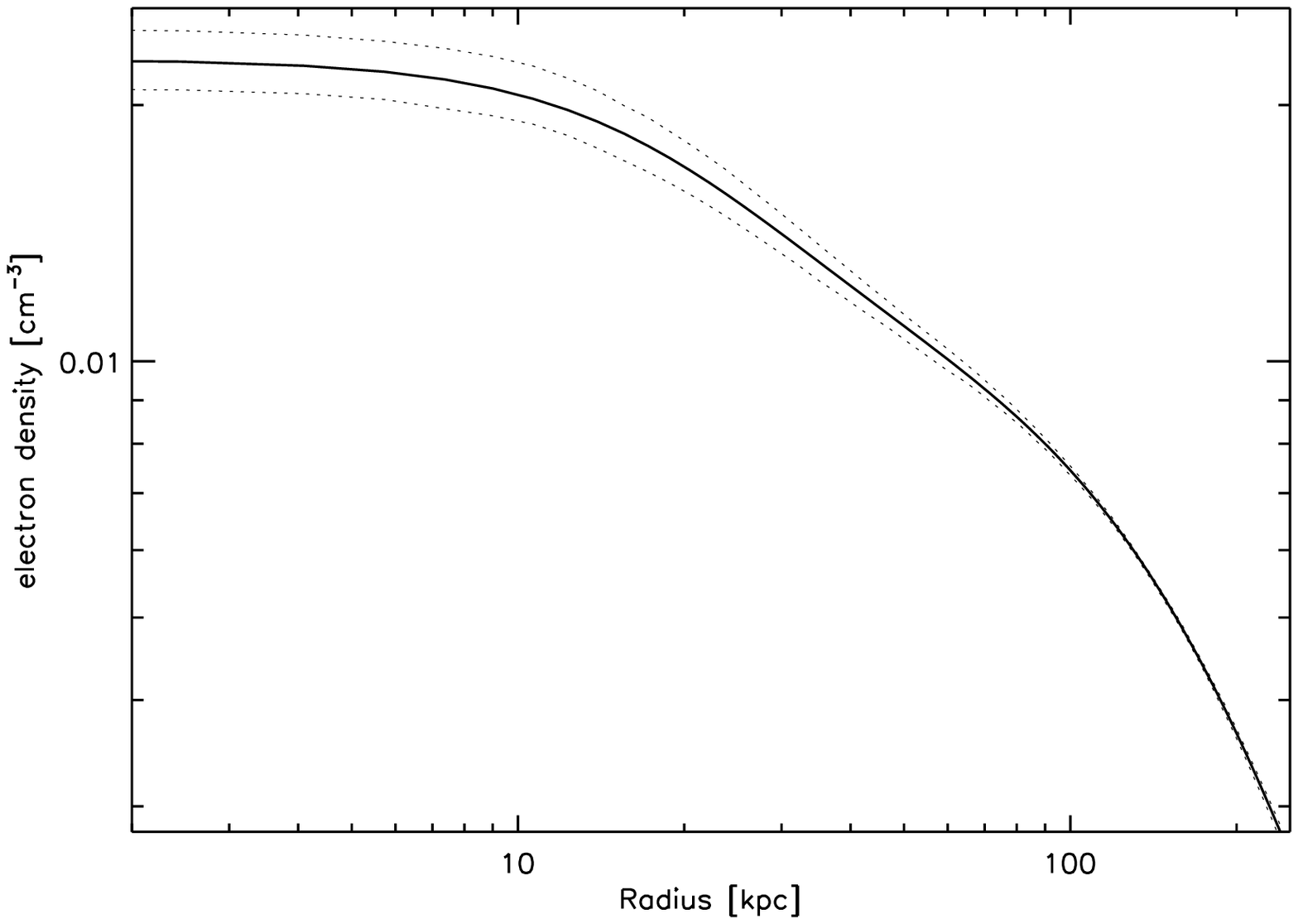}}
\caption{In the left panel the black crosses show the surface brightness data of MOS2 centred at the X-ray peak. 
The best-fit narrow and wide $\beta$ components are shown with a dashed red line and a dotted green line,
respectively. The blue line is the sum of the two $\beta$-components. The models are extrapolated to
r$<$0.2 arcmin (= 7 kpc). The derived best-fit gas density profile in shown in the right panel 
as a thick line, together with the uncertainties due to the statistical errors of the parameters at 1$\sigma$ 
level.
\label{profile.fig}}
\end{figure*}

We extracted a surface brightness profile in co-centric annuli of 12 arcsec width
centred at the X-ray peak location (see Fig. \ref{profile.fig}). 
The profile is not well fitted with a single-$\beta$ model (Cavaliere \& Fusco-Femiano, 1976) and thus we used the double-$\beta$ model, which describes the
data well in the radial range 12'' -- 7' with core radii 
r$_{core,1}$ = 0.7$\pm$0.1 arcmin and r$_{core,2}$ = 4.9$\pm$0.2 arcmin in the two components (narrow and broad 
$\beta$-components), while the common $\beta$-parameter obtains a value of 0.74$\pm$0.02.

\subsection{Cool core}
\label{cd_cooling}

A natural explanation of the emission in the narrow $\beta$-component is the cooling of the intra-cluster gas.
Suzaku analysis of Ophiuchus suggested that Ophiuchus has a cool core (Fujita et al., 2008).
The typical cooling radius of 70 kpc ( = 2 arcmin) found in a cluster sample by the Chandra satellite 
(Vikhlinin et al., 2006) is also consistent with the extent of the central excess.

To examine in detail the central emission we accumulated a PN spectrum within a r=30'' radius circle around the X-ray peak
(for more details about the spectral fitting, see Section \ref{spectralfitting}). 
The spectrum was not acceptably fitted with a single temperature model, and we thus experimented with a two-temperature
model. Since the metal abundance of the cool component was poorly constrained, we fixed it to
1.0 in Solar units. This model yielded an acceptable solution whereby $\sim$90\% of the total emission  
originates from (hot) gas with T=6.7$\pm$0.4 keV, while the rest is due to a (cool) component with T=1.7$\pm$0.1 keV.

The temperature of the hot component is $\sim$30\% lower than that outside 
the central 1 arcmin region (see below). Such a temperature drop is consistent with those found in 
the Chandra sample of clusters (Vikhlinin et al., 2006). 
The emission measure of the hot component yields a central proton density of $\sim 2 \times 10^{-2}$ cm$^{-3}$
(see Fig. \ref{profile.fig})
corresponding to a cooling time of $3 \times 10^9$ yr. 
This argues against a strong recent merger.

However the cooling radius in Ophiuchus is quite small 
($\sim$30 kpc) compared to other clusters. Also, the proton density decreases only by 40\% between r=0 and r=1.0 arcmin
which increases the cooling time by a factor of 1.7, i.e. to $5 \times 10^9$ yr at r=1 arcmin. The rapid drop in the 
density (and the rapid increase in cooling time) begins at r=3 arcmin where one would assume the cooling to end.
Note that the radius 3 arcmin (= 100 kpc) is better consistent with the cooling radii in the Chandra sample. 
Thus it appears that the 1--3 arcmin temperatures are too high for the typical cool core cluster.

\subsection{CD galaxy}
As noted in Perez-Torres et al. (2009), the X-ray peak location is consistent with that of an elliptical cD galaxy 
2MASX J17122774-2322108 (Hasegawa et al., 2000). 
The extent of the central excess on top of the broad $\beta$-component ($\sim$ 2 arcmin) corresponds to 
70 kpc at the distance of Ophiuchus. This is consistent with the maximal extent of X-ray haloes of cD galaxies
(e.g. Matsushita, 2001). 
Also, in the innermost 12'' the surface brightness data exceeds the model by 30\%, indicative of a 
central point source which may be the nucleus of the cD galaxy. Thus the spatial distribution of the X-ray brightness
is consistent with a central cD galaxy.

The temperature and the 0.2--2.0 keV band luminosity (1.8 $\times 10^{42}$ erg s$^{-1}$) of the cool component 
in the above fit (Section \ref{cd_cooling}) are consistent with those in several cD galaxies studied in e.g. Matsushita (2001).
The surface brightness profile derived above shows that the flux in the narrow $\beta$-component in the central 30'' region
exceeds that of the broad $\beta$-component.
Since the cool component in the above fit constitutes only 10\% of the central flux, it does not explain the central 
brightness excess as modelled with the narrow $\beta$-model. 
Rather, the flux of the cool component is consistent with the point-source-like excess flux in the central 12'' region on
top of the double-$\beta$ model prediction (see Fig \ref{profile.fig}). We thus attribute the cD galaxy emission with the 
low temperature point source at the cluster centre.

\begin{figure*}[t]
\hbox{
\hbox{
\includegraphics[width=8cm,angle=0]{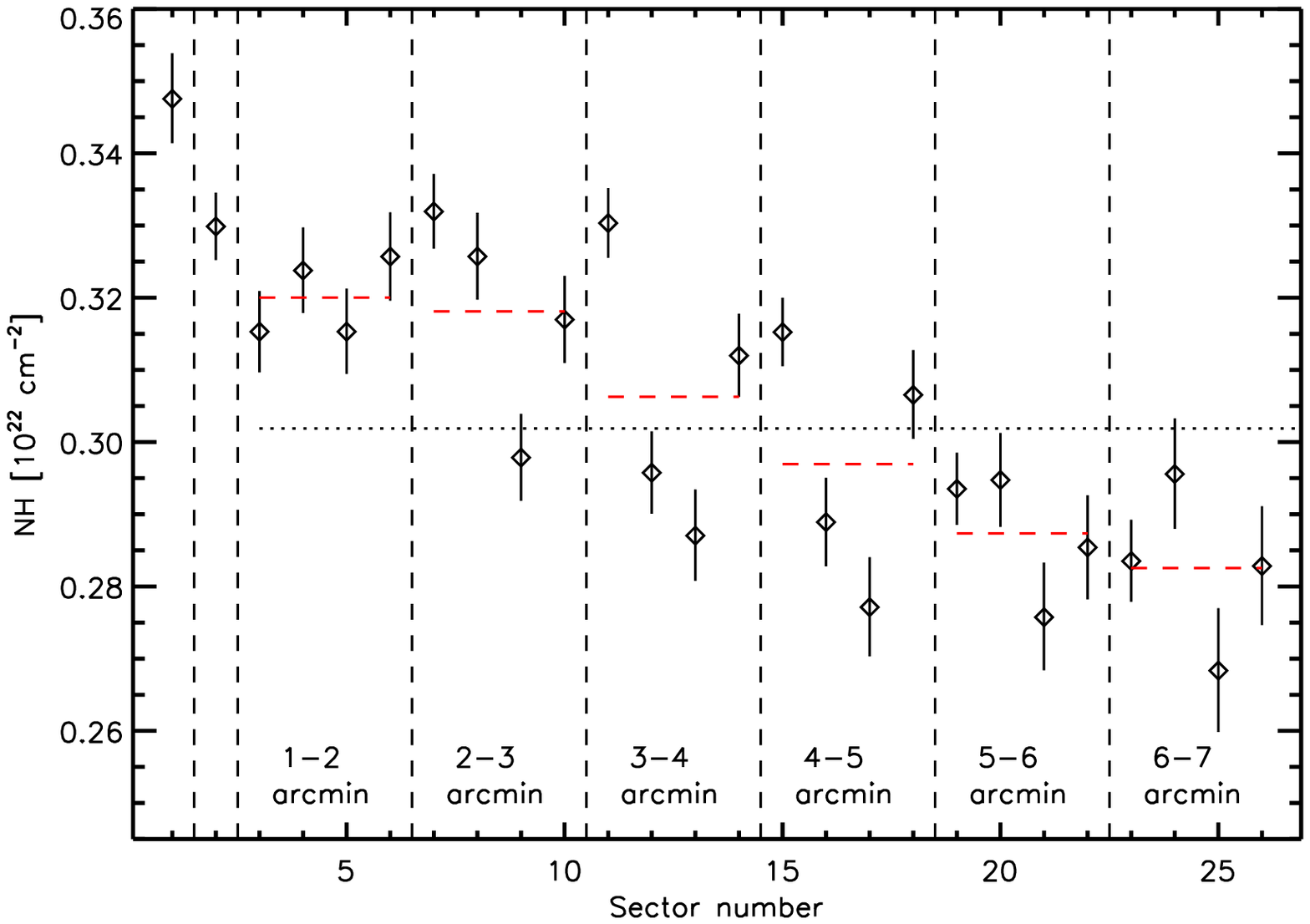}
\includegraphics[width=8cm,angle=0]{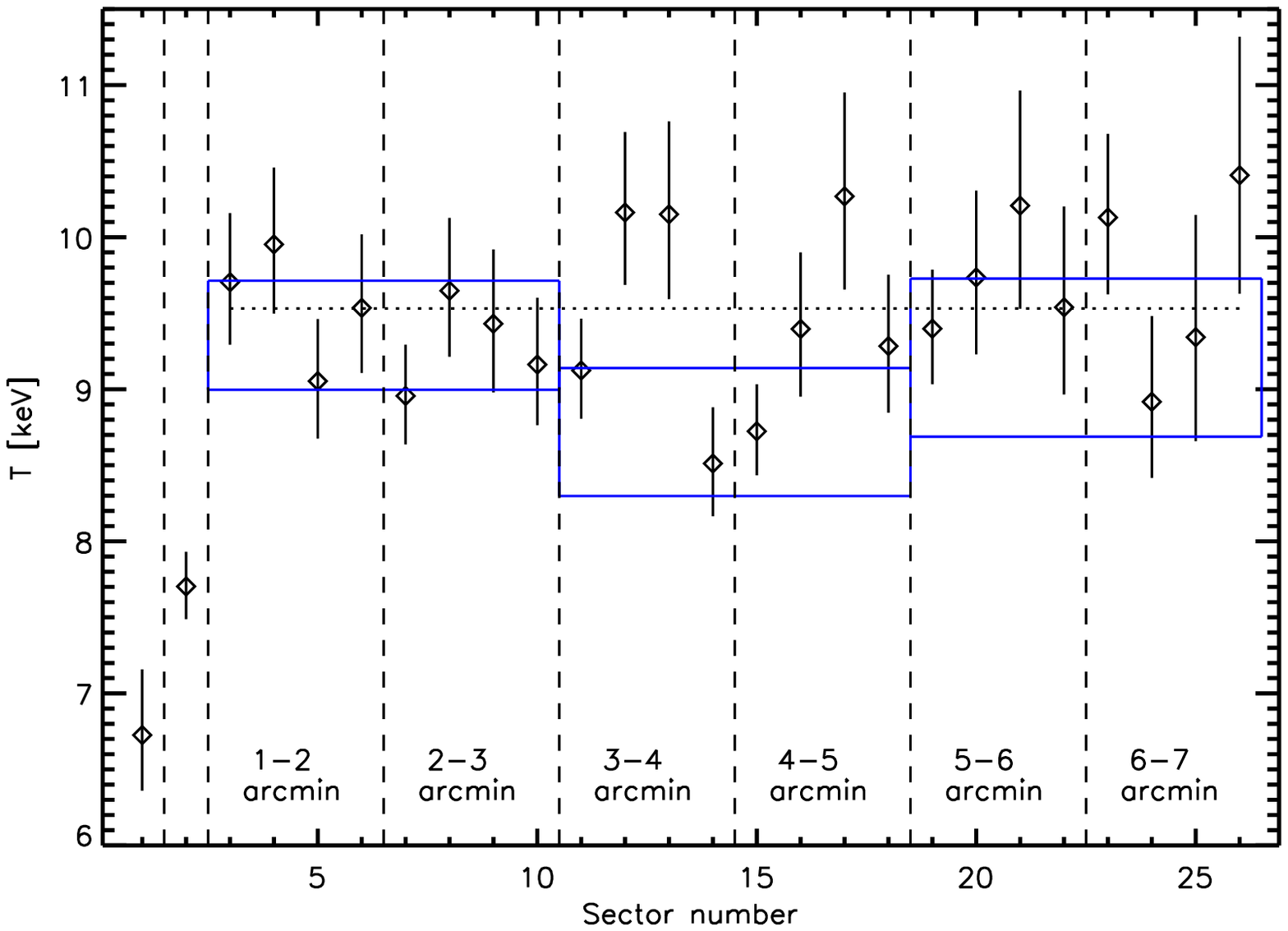}}}
\caption{The best-fit values and statistical uncertainties (diamonds and solid lines)
for the free NH (left panel) and the temperature (right panel). 
The blue boxes show the temperatures derived using the Fe XXV/XXVI line ratio.
The vertical dashed lines separate the different radial regions. 
For a given annulus, the sectors are numbered clock-wise from the North-West, except for ``sectors'' 1 and 2, which 
refer to the r=0.5 arcmin circle and r=0.5--1.0 arcmin annulus. 
The horizontal dashed and dotted lines show the azimuthal average values.  
\label{sectors.fig}}
\end{figure*}

\section{XMM-Newton spectral analysis}
\label{spectralfitting}
In order to examine the spatial distribution of the temperature of the intra-cluster gas we performed spatially
resolved spectroscopy using PN data. We divided the central r=7 arcmin region into annuli of 0.5--1 arcmin width and 
further divided the annuli at r=1--7 arcmin into four sectors. We centred the annuli on the surface brightness 
peak.
We extracted spectra in these regions and 
fitted the 0.5--7.4 keV band data with a single-temperature MEKAL model, except in the central r=0.5 arcmin region, 
where an additional thermal component is required (as discussed in Section \ref{cd_cooling}). 
We fitted the spectra using XSPEC version 12.5.0s.
We adopted the solar abundance table of Anders \& Grevesse N. (1989).
We note that we assume in this modelling that the non-thermal component 
does not affect significantly the derived spatial distribution of the thermal properties, i.e. that the 
non-thermal component contributes insignificantly in the PN band compared to the thermal component or 
that the spatial distribution of the non-thermal component is azimuthally symmetric. 

\subsection{NH}
\label{nh}
We attempted to model the spectra including a WABS absorption model which uses the photo-electric absorption cross 
sections of Morrison and McCammon (1983) and the relative element abundances of Anders and Ebihara (1982).
However, we were unable to model the spectra acceptably when using the hydrogen column density NH = 0.2 $\times 10^{22}$ 
cm$^{-2}$ as given by the 21 cm radio measurements of neutral hydrogen (HI) (Kalberla et al., 2005).
Also, the best-fit temperatures are unrealistically high (see Table \ref{sectors_t_tab}).
Neither the absorption model PHABS nor the different tables for the cross section or relative abundances in XSPEC 
improved the fits.

While the HI-based NH value (Kalberla et al., 2005) varies in the range 0.16--0.23  $\times 10^{22}$ cm$^{-2}$ in the 
fields centred within  1$^{\circ}$ of the Ophiuchus centre, the PN data require a significantly larger amount of 
absorption (NH = 0.3--0.4  $\times 10^{22}$ cm$^{-2}$, see Table \ref{sectors_t_tab}).
Similar results were reported in the Suzaku analysis of Ophiuchus (Fujita et al., 2008). 
We also found that the modelling of the soft X-ray sky background using ROSAT All Sky Survey data close to Ophiuchus 
yielded a higher value NH = 0.46 $\times 10^{22}$ cm$^{-2}$ (Section \ref{back}).
Note that the NH values of Kalberla et al. (2005) are derived under the assumption that the HI 21-cm line is optically
thin, which may be violated at latitudes within 10$^{\circ}$ of the Galactic equator and thus the values may be 
underestimated.\footnote{http://www.astro.uni-bonn.de/$\sim$webaiub/english/tools\_labsurvey.php}

\begin{table*}
\caption[]{Single-temperature fits}
\label{sectors_t_tab}
\centering
\renewcommand{\footnoterule}{}  
\begin{tabular}{ccc|ccc}
\hline\hline
\multicolumn{3}{c}{NH fixed$^{a}$}       & \multicolumn{3}{c}{NH free$^{b}$}\\                
region$^{c}$       &  T [keV] & $\chi^2$/d.o.f & NH [$10^{22}$ cm$^{-2}$]              & T [keV]         & $\chi^2$/d.o.f \\
\hline
0$'$-0.5$'$$^{d}$       &  79.9  & 987.0/284 &  0.35[0.34--0.35] & 6.7[6.4--7.2] & 209.0/283    \\
0.5$'$-1$'$  & 23.0     &  1361.2/447    & 0.33[0.33--0.33]  & 7.7[7.5--7.9]   &  373.0/446     \\
1$'$-2$'$ nw & 33.5     &  740.9/332     & 0.32[0.31--0.32]  & 9.7[9.3--10.2]  & 242.5/331      \\
1$'$-2$'$ sw & 38.9     &  780.3/312     & 0.32[0.32--0.33]  & 10.0[9.5--10.5] & 251.0/311      \\
1$'$-2$'$ se & 27.5     &  651.7/292     & 0.32[0.31--0.32]  & 9.1[8.7--9.5]   & 202.2/291      \\
1$'$-2$'$ ne & 36.7     &  762.7/301     & 0.33[0.32--0.33]  & 9.5[9.1--10.0]  & 248.0/300      \\
2$'$-3$'$ nw & 31.9     &  1128.6/417    & 0.33[0.33--0.34]  & 9.0[8.6--9.3]   & 301.1/416      \\
2$'$-3$'$ sw & 35.4     &  810.2/317     & 0.33[0.32--0.33]  & 9.6[9.2--10.1]  & 253.4/316      \\
2$'$-3$'$ se & 23.0     &  546.3/295     & 0.30[0.29--0.30]  & 9.4[9.0--9.9]   & 227.0/294      \\
2$'$-3$'$ ne & 30.7     &  703.7/298     & 0.32[0.31--0.32]  & 9.2[8.8--9.6]   & 250.6/297      \\
3$'$-4$'$ nw & 34.5     &  1255.8/465    & 0.33[0.33--0.34]  & 9.1[8.8--9.5]   & 362.3/464      \\
3$'$-4$'$ sw & 26.8     &  574.4/306     & 0.30[0.29--0.30]  & 10.2[9.7--10.7] & 249.3/305      \\
3$'$-4$'$ se & 24.3     &  427.3/268       & 0.29[0.28--0.29]  & 10.1[9.6--10.8] & 210.8/267      \\
3$'$-4$'$ ne & 24.2     &  704.2/312     & 0.31[0.31--0.32]  & 8.5[8.2--8.9]   & 257.9/311      \\
4$'$-5$'$ nw & 24.5     &  1104.2/457    & 0.32[0.31--0.32]  & 8.7[8.4--9.0]   & 374.5/456      \\
4$'$-5$'$ sw & 21.0     &  442.8/279     & 0.29[0.28--0.30]  & 9.4[9.0--9.9]   & 201.4/278      \\
4$'$-5$'$ se & 21.0     &  332.3/250     & 0.28[0.27--0.28]  & 10.3[9.7--11.0] & 186.6/249      \\
4$'$-5$'$ ne & 25.6     &  571.2/284     & 0.31[0.30--0.31]  & 9.3[8.8--9.8]   & 207.3/283      \\
5$'$-6$'$ nw & 21.6     &  798.4/409     & 0.29[0.29--0.30]  & 9.4[9.0--9.8]   & 383.5/408      \\
5$'$-6$'$ sw & 24.4     &  448.7/264     & 0.29[0.29--0.30]  & 9.7[9.2--10.3]  & 199.3/263      \\
5$'$-6$'$ se & 21.5     &  296.8/215     & 0.28[0.27--0.28]  & 10.2[9.5--11.0] & 178.6/214      \\
5$'$-6$'$ ne & 21.7     &  341.2/241     & 0.29[0.28--0.29]  & 9.5[9.0--10.2]  & 181.2/240      \\
6$'$-7$'$ nw & 22.9     &  492.7/309     & 0.28[0.28--0.29]  & 10.1[9.6--10.7] & 243.7/308      \\
6$'$-7$'$ sw & 20.8     &  370.1/216     & 0.30[0.29--0.30]  & 8.9[8.4--9.5]   & 187.6/215      \\
6$'$-7$'$ se & 17.7     &  181.9/173     & 0.27[0.26--0.28]  & 9.3[8.7--10.1]  & 112.3/172      \\
6$'$-7$'$ ne & 24.9     &  241.6/203     & 0.28[0.27--0.29]  & 10.4[9.6--11.3] & 122.4/202      \\
\hline
             &          &                &                   &                 &                \\
\multicolumn{6}{l}{$^{a}$: NH is fixed to the 21 cm measurement (= 0.2 $\times 10^{22}$ cm$^{-2}$)} \\
\multicolumn{6}{l}{$^{b}$: NH is a free parameter}   \\
\multicolumn{6}{l}{$^{c}$: The values indicate the inner and outer radii while the letters indicate the direction of}\\
\multicolumn{6}{l}{the spectrum extraction sector: North-West (nw), South-West (sw), South-East (se),}\\ 
\multicolumn{6}{l}{and North-East (ne)} \\
\multicolumn{6}{l}{$^{d}$: The model includes a component for the CD galaxy}\\
\end{tabular}
\end{table*}

At low Galactic latitudes molecular material may yield significant absorption, and yet remain undetected 
in radio surveys since it does not contribute to the 21 cm line.
This is quite probable for the Ophiuchus region, which is a well known site of star formation.  
The absorption in our energy band of 0.6--7.4 keV is dominated by heavier elements than hydrogen.
We use the molecular hydrogen (H$_2$) as a tracer of molecular metals in the following.
Unfortunately a direct measurement of the H$_2$ column density towards the Ophiuchus cluster is not reported in the 
literature.
The measurements of the interstellar dust towards Ophiuchus can be used to estimate the total NH, since 
the dust is physically connected to the neutral and molecular hydrogen.  
Schlegel et al. (1998) mapped the density of dust in the Galaxy using IRAS 100 $\mu$m and DIRBE FIR data.
Using the background galaxies they derived the dust extinction and yielded a reddening value of  
E(B-V) = 0.585 mag in the direction of Ophiuchus cluster.

The relation between E(B-V) and total HI+H$_2$ density was calibrated using 
the L-$\alpha$ absorption measurements of stars with Copernicus-satellite (Bohlin et al, 1978).
The relation with the above estimate for the reddening in Ophiuchus yields a total NH of 0.34 $\times 10^{22}$ cm$^{-2}$,
much higher than the 21cm value (0.2 $\times 10^{22}$ cm$^{-2}$), indicating that H$_2$ density is substantial, 
35\% of that of HI. 
The value of total NH is consistent with the highest X-ray absorption value in our work (see Fig. \ref{sectors.fig}).
Assuming that the metal to hydrogen abundance ratios in the molecular material are approximately Solar, the above 
consistence means that the molecular metal absorption is a viable explanation for the difference between the NH value 
obtained from X-ray absorption and radio 21cm emission.
In the following we thus account for the excess absorption by allowing NH to be a free parameter of the WABS model.

\subsection{Temperature distribution}
\label{tempe}
Fortunately we can also constrain the temperatures without the effect of NH uncertainties, 
utilising the monotonically decreasing Fe XXV/XXVI emission line ratio 
with temperature.
In practise, we fitted the data with a MEKAL model in a narrow band (6.0--7.4 keV) containing 
both lines (redshifted to $\sim$6.5 and $\sim$6.8 keV at the distance of Ophiuchus), where the absorption is negligible. 
The drawback is that the number of photons in the above regions is too small in this narrow band for a statistically 
meaningful analysis. We thus used 1--3--5--7 arcmin annuli to examine the mean temperature with the Fe lines. 

Most of the thus derived temperatures agree
with those derived with the full band and free NH (see Fig. \ref{sectors.fig}).
This means that the NH enhangement towards the cluster centre is real and probably due to a chance superposition of the 
Ophiuchus center and a denser part of a Galactic molecular matter clump.
The temperatures in sectors 12, 13 and 17 disagree with the azimuthal average, but only by $\sim$2$\sigma$.
Thus, we do not detect very significant azimuthal temperature variations in the central 7 arcmin region of Ophiuchus.
This implies that the central region of Ophiuchus is not disturbed by a recent major merger.

Utilising the azimuthal symmetry, we further examined the radial behaviour of the temperature 
in concentric annuli of 0.5 arcmin width and
fitted the spectra with NH as a free parameter with a single-temperature MEKAL component
(keeping the cool component in the centre as found above).  
At 1--7 arcmin the profile is very flat, consistent with the value obtained from a fit to a single spectrum of the data 
in the radial range 1--7 arcmin (9.1$\pm$0.1 keV) (see Fig. \ref{sectors.fig}).
The central temperature drop (discussed above) 
is accompanied by the central increase of the metal abundance up to 0.6 Solar from the constant $\sim$0.3 Solar
at r = 2--7 arcmin, as observed in cool core clusters (e.g. Baldi et al., 2007).

\begin{figure*}
\hbox{
\includegraphics[width=9cm,angle=0]{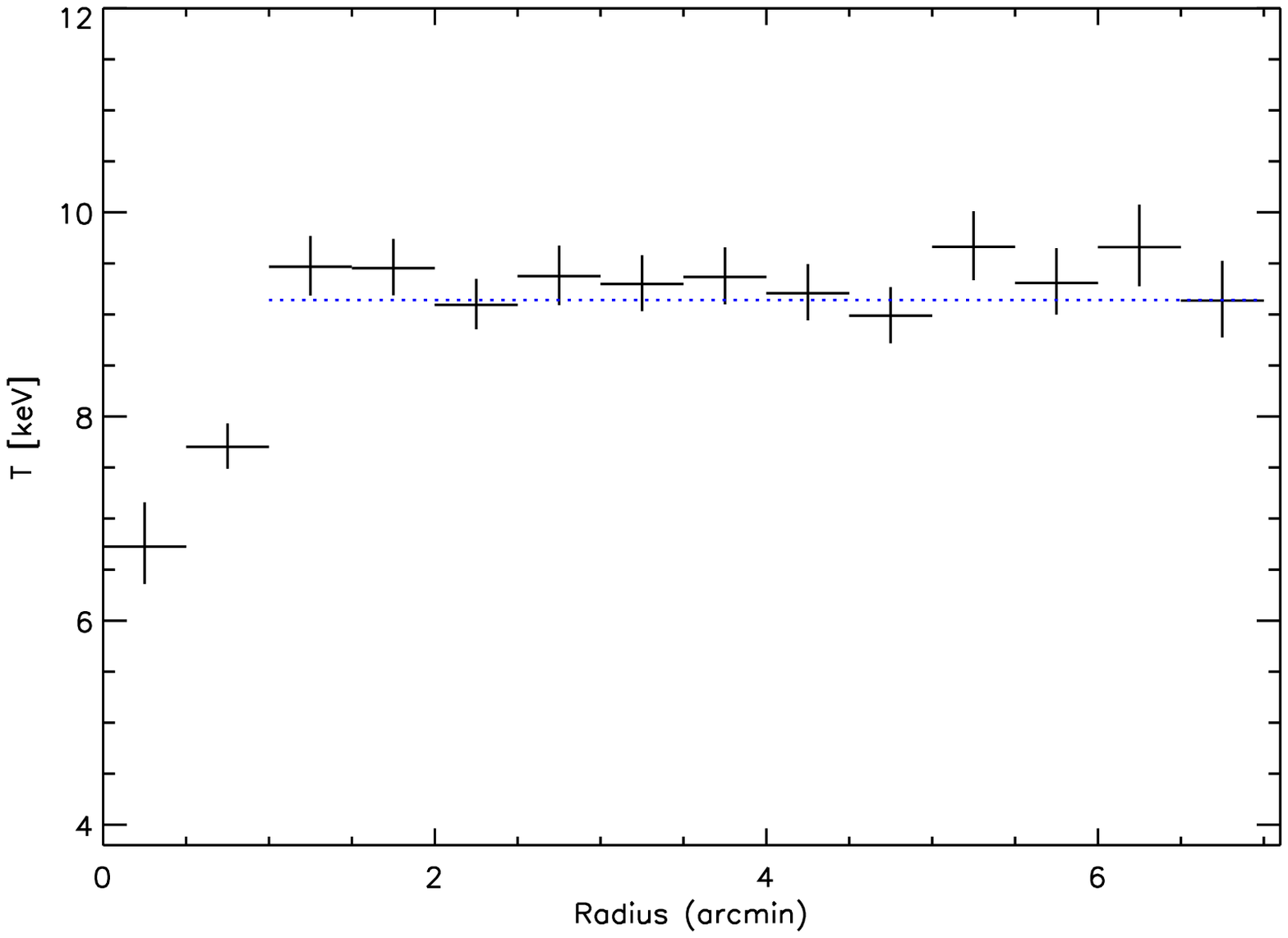}
\includegraphics[width=9cm,angle=0]{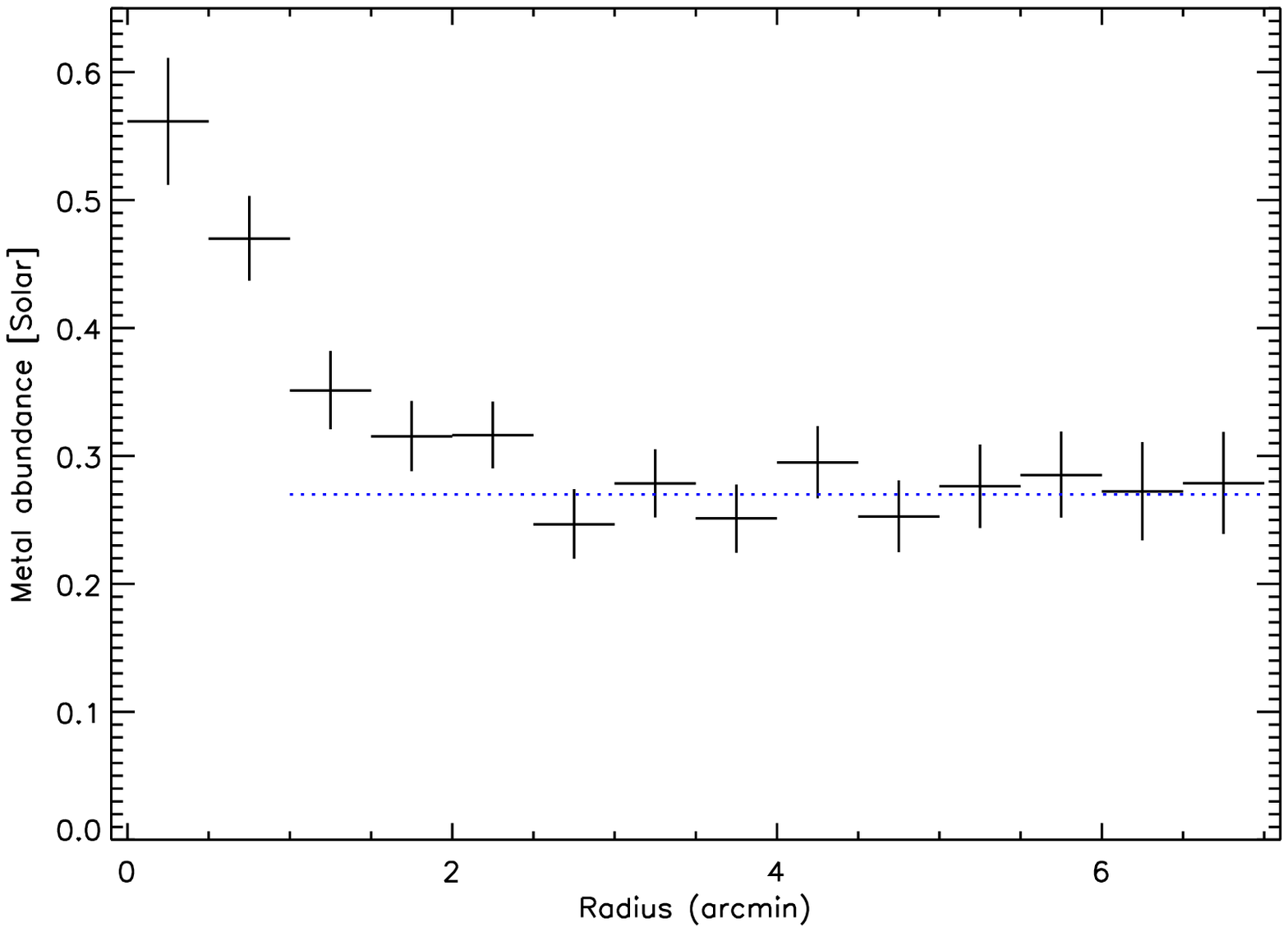}
}
\caption{The best-fit values and 1$\sigma$ uncertainties of the temperature (left panel) and the metal abundance 
(right panel) profiles of Ophiuchus. The horizontal lines show the values obtained from a fit to a spectrum accumulated
in a 1--7 arcmin annulus.
\label{gasphysics.fig}}
\end{figure*}

\subsection{Thermal model of the central 7 arcmin region}
\label{3tmodel}
Using the best-fit double-$\beta$ model for the flux distribution we estimated that 90\% of the total flux
in the central 7 arcmin region flux originates from the 1--7 arcmin region. This is consistent with the emission measures
obtained from the spectral fits to the spectra accumulated in the 0--1--7 arcmin regions.
Thus most of the emission is due to isothermal gas which simplifies the total emission modelling.
Our final XMM-Newton based thermal model for the emission in the central 7 arcmin region of Ophiuchus consists of 
three thermal components. One of the components models the emission from the cD galaxy (as derived above), 
i.e. with temperature and metal abundances 
T$_{1}$ = 1.7 keV, abund$_1$ = 1.0 Solar. The second component describes the cool gas in the core, obtained 
with a fit to 0--1.0 arcmin data while keeping the above cD component fixed. This component has parameters
T$_{2}$ = 7.3 keV, abund$_2$ = 0.50 Solar. 
The third component is for the isothermal gas in the 1--7 arcmin region with 
T$_{3}$ = 9.1 keV, abund$_3$ = 0.27 Solar.

\begin{figure}
\includegraphics[width=9.5cm,angle=0]{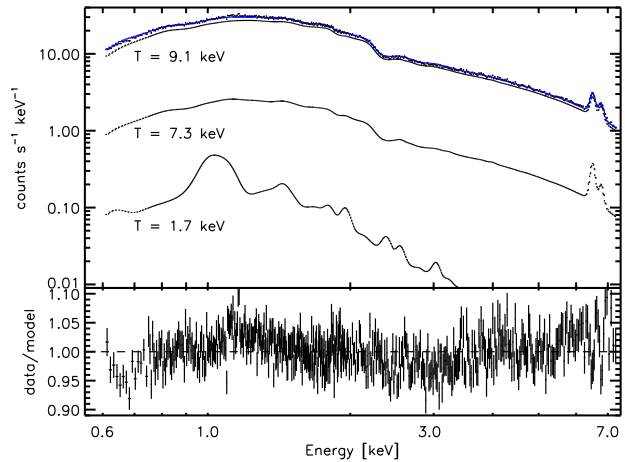}
\caption{In the upper panel the crosses show the PN data and statistical uncertainties from the central r=7 arcmin
region. The blue line shows the best-fit 3T model. The emission model components are also shown separately as solid lines.
The lower panel shows that ratio of the data and the model prediction.
\label{pn_3t.fig}}
\end{figure}

Comparison of this model ($\equiv$ 3T model, see Fig. \ref{pn_3t.fig})) with the spectrum of the full 7 arcmin region 
shows that there are some problems. 
There appears an un-modeled emission feature at 1.2 keV which cannot be explained by 
background uncertainties (see Molendi \& Gastaldello, 2009, for a discussion of a similar feature in case of 
Perseus cluster).
For the purpose of this work, we noted 
that excluding the 1.1--1.3 keV band from  our spectral fits did not produce a significant change to the results. 
Also, there are systematically increasing residuals in the energy range 0.6--1.0 keV at the $\sim$ 5\% level. 
Furthermore, the Fe XXV and XXVI lines are not well modeled 
(see below). We note that the large number of photons ($\sim 10^{6}$) in the Ophiuchus spectrum 
renders the statistical uncertainties small ($\sim$ 2--10\% in the 0.6--7.4 keV band) even with the original binning. 
It is possible that remaining calibration uncertainties of the total efficiency of PN at this level become important due to 
the high statistical precision of the data.

Note that we have assumed here that the non-thermal component does not affect the fits. Thus, when including the INTEGRAL 
data and a model for the non-thermal emission, we will examine and discuss the modifications to the thermal modelling.

\section{Joint XMM-Newton and INTEGRAL analysis}
With the thermal emission modeled, we then proceeded to a joint XMM-Newton -- INTEGRAL modelling of the total emission in 
a very wide band, 0.6--140 keV, of the central circular region within a radius of 7 arcmin. 

\subsection{PN/ISGRI cross-calibration}
\label{pn_isgri}
When comparing the different instruments, we noted that the central 7 arcmin area covered by PN is obscured by 8\% due 
to CCD gaps and bad pixels.

Given that the Ophiuchus cluster shows an angular extent larger than the PSF of IBIS/ISGRI, for a correct 
cross-calibration between the different instruments it is necessary to know the fraction of the flux extracted by ISGRI 
compared to the total flux in a radius of 7 arcmin used for the PN spectral extraction (see below). 
To compute this number, we 
computed the shadow pattern cast by the source using the surface-brightness profile as measured by XMM/MOS2 (see 
Eckert et al., 2007 and Renaud et al., 2006, for the details), and simulated raw ISGRI detector images with Poissonian 
background in each energy band. Then we extracted the spectra in the standard way and computed the total flux extracted 
by the software. As a result, we found that ISGRI extracts 84\% of the total injected flux.

We assumed that the $\sim$6\% difference in the covered area between PN and INTEGRAL does not significantly change the 
spectral shape and we thus multiplied the model normalisations with the covering fractions in the fitting models 
so that both obtained best-fit model normalisations correspond to a full unobscured r=7 arcmin circular region. 
We additionally allowed 10\% variation between the above PN--ISGRI normalisation factor due to systematic uncertainties 
in the PN/ISGRI cross-calibration.

\subsection{Excess emission}

\begin{figure}
\includegraphics[width=9cm,angle=0]{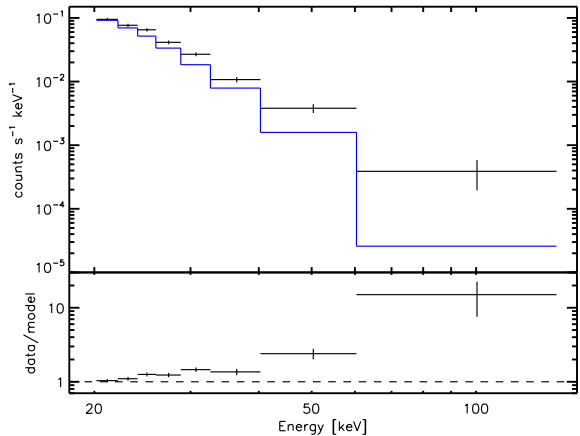}
\caption{In the upper panel the crosses show the ISGRI data, while the blue line shows the best-fit 3T model 
extrapolation from the PN analysis. The lower panel shows the ratio of data and model prediction. 
\label{isgri_3t.fig}}
\end{figure}

The prediction of the 3T thermal model derived with XMM-Newton data in Section \ref{3tmodel} fails to reproduce the 
ISGRI data at energies above 25 keV (see Fig. \ref{isgri_3t.fig}).
Combining the effect of the statistical uncertainties of the thermal model parameters 
and the statistical uncertainties of the ISGRI data we found that the 
ISGRI data exceeds the model prediction by 5.7$\sigma$ in the 20--140 keV band, 
by a maximum difference of a factor of $\sim$10 at 100 keV, similar to the results in Eckert et al. (2008).

To quantify the effect of the excess emission observed by ISGRI we fitted the ISGRI data in the 20--140 keV band 
using a single-temperature MEKAL model (abundance fixed to 0.3 Solar). The resulting temperature value is 
T=11.6$\pm$0.6, i.e. much higher than the hot component measured with PN (9.1$\pm$0.1 keV).
The best-fit thermal model to 20--80 keV band of ISGRI data in the central 7 arcmin region (scaled to the 
full unobscured region based on simulations, see Section \ref{pn_isgri}) yields a flux of 
$\sim$4.5 $\times 10^{-11}$ erg s$^{-1}$ cm$^{-2}$ in this band. 
Note that this model does not fit well the data at the highest energies.
Since the detection of radio emission proved the existence of relativistic electrons in Ophiuchus (Govoni et al., 2009)
we assume in the following that the excess emission found here is non-thermal.

Because of the broad PSF of IBIS/ISGRI (12 arcmin FWHM), we cannot exclude the possibility that the hard X-ray 
excess detected by INTEGRAL would be due to a highly-obscured Seyfert nucleus, which would be hidden in the soft X-ray
band (1--10) keV due to high NH. However, the population studies of AGN in hard X-rays by INTEGRAL (Beckmann et al. 
06, Paltani et al. 08) and Swift (Tueller et al. 08) have shown that there is no evidence for such a Compton thick AGN 
population which is not detected in the soft X-rays. In order to examine this issue quantitatively, we used ISGRI
data of Ophiuchus to examine the possible time variability produced by AGN in the field. 
We extracted spectra from periods 
March 28, 2003 - August 23, 2004; August 26, 2004 -- September 22, 2005 and September 22, 2005 -- September 14, 2007.
The exposure times are 1.1, 0.8 and 1.2 Msec. The resulting spectra are consistent within the statistical 
uncertainties: The best-fit MEKAL spectra yield consistent temperatures 
(12.2$\pm$1.0 keV,  12.3$\pm$1.3 keV, 10.6$\pm$0.9 keV) and 
20--80 keV band fluxes\footnote{The fluxes are scaled to the full r=7 arcmin circular region based on the simulations in 
Section \ref{pn_isgri}.} (4.7$\pm$0.4 $\times 10^{-11}$ erg s$^{-1}$ cm$^{-2}$, 
4.5$\pm$0.5, $\times 10^{-11}$ erg s$^{-1}$ cm$^{-2}$ and 4.3$\pm$0.3 $\times 10^{-11}$ erg s$^{-1}$ cm$^{-2}$), 
respectively for the different periods.
Thus, these data indicate no variability in the Ophiuchus hard spectrum during the 4 years of observations
which argues against a significant AGN contribution.

\subsection{Modelling the inverse compton scattering of CMB photons}
The usual mechanism for explaining the hard X-ray excess in clusters is the inverse compton scattering (IC) of the cosmic
microwave background (CMB) photons by the relativistic cluster electrons (e.g. Sarazin, 1988). 
We examined this scenario by performing spectroscopy to PN and ISGRI data in the following.
If the same population of relativistic electrons produces both the synchrotron emission and the IC,
the spectral indices of the two power-law spectra should be equal. Unfortunately this test is not yet possible, 
because the radio halo in Ophiuchus has been significantly detected only at one wavelength. 
While keeping the thermal model fixed to the above 3T model,
we added a simple power-law and its variants i.e. with two photon indices on both sides
of a break energy (broken power-law) or with an exponential high energy cut-off (cutoff power-law) to the spectral model. 
When deriving the thermal model above we assumed that the non-thermal emission is negligible at energies below 10 keV,
compared to the thermal emission. This requires a very hard simple power-law and such a model cannot reproduce 
the ISGRI data well. Using the broken power-law and cutoff power-law models the 
photon index becomes negative at energies below 10 keV. 
Since such models are unphysical, we then allowed in the following the  
hot component of the 3T model (the MEKAL model with T$\sim$ 9 keV) to be modified by the addition of the non-thermal component. 
We kept the central cooler components fixed because the thermal emission is so bright in the centre that it probably 
dominates the emission.

The resulting best-fit model has a steep power-law, with $\alpha_{ph}$ = 3.4$\pm$0.2.
The NH obtains a very high value, 0.44$\pm$0.02 $\times 10^{22}$ cm$^{-2}$, much higher 
than that derived from the IRAS and Copernicus data (see Section \ref{nh}).
Also, the model underpredicts the ISGRI data significantly at E$>$40 keV.
The high level of background emission results in large statistical uncertainties in the ISGRI data 
($\sim$50\% at 100 keV),
and thus the statistical weight of ISGRI data is negligible compared to that of PN. This is demonstated by the fact that 
when excluding the ISGRI data from the fit the power-law properties are consistent with those in the joint PN+ISGRI fit 
above. 
It is unfortunate that the statistical weight of the data is smallest, where the signal is relatively strongest:
The emission of the thermal model decreases exponentially at energies above 10 keV, while the non-thermal component
is assumed to have a power-law spectrum. Thus a larger fraction of the ISGRI data consists of non-thermal emission, 
compared to PN.
The PN data dominate the fit and thus small uncertainties of the modelling of the thermal 
emission due to problems of e.g. NH or the PN calibration (see Section \ref{spectralfitting})   
may weigh more than the real ISGRI data. 

\begin{figure*}
\hbox{
\includegraphics[width=9cm,angle=0]{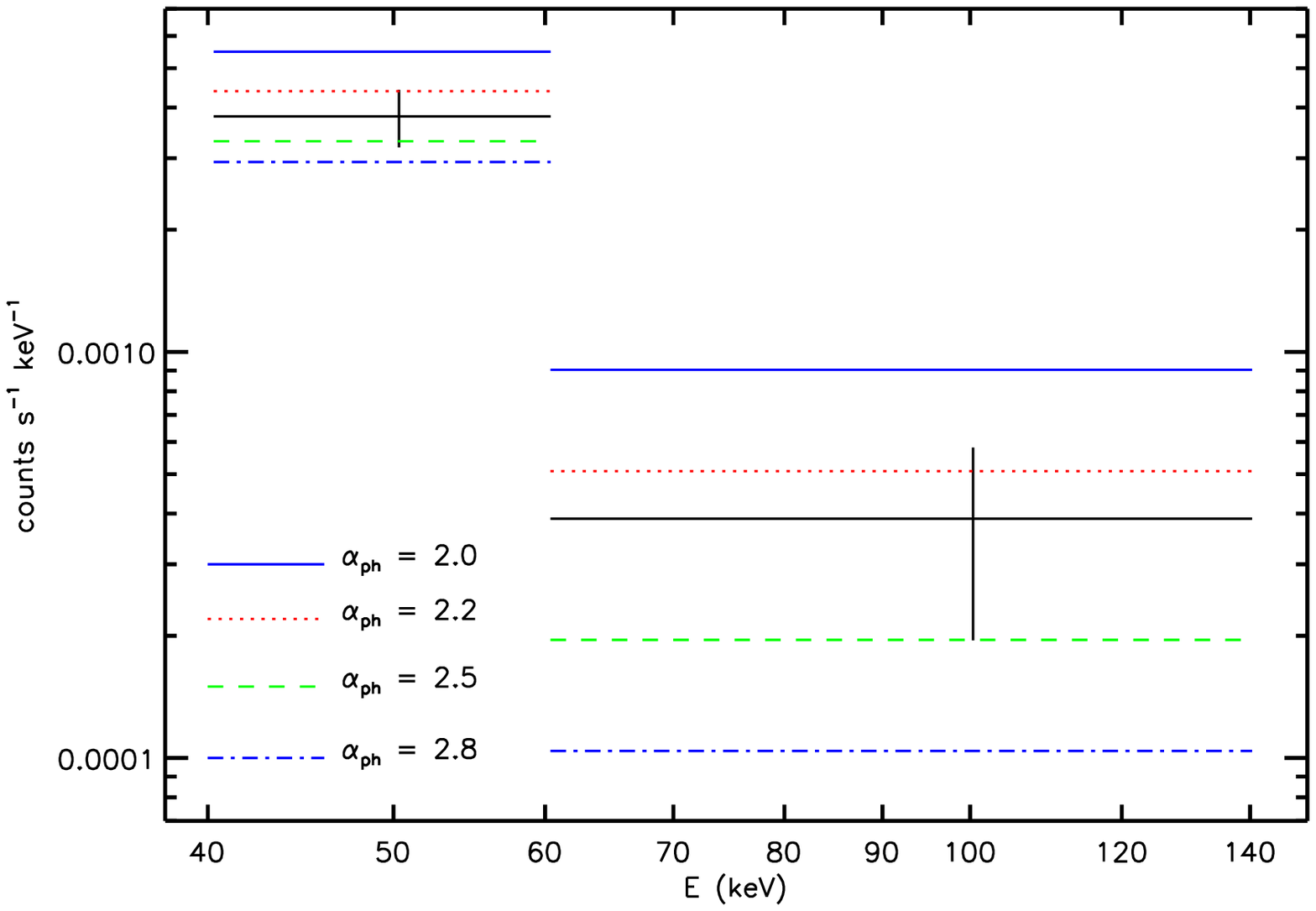}
\includegraphics[width=9cm,angle=0]{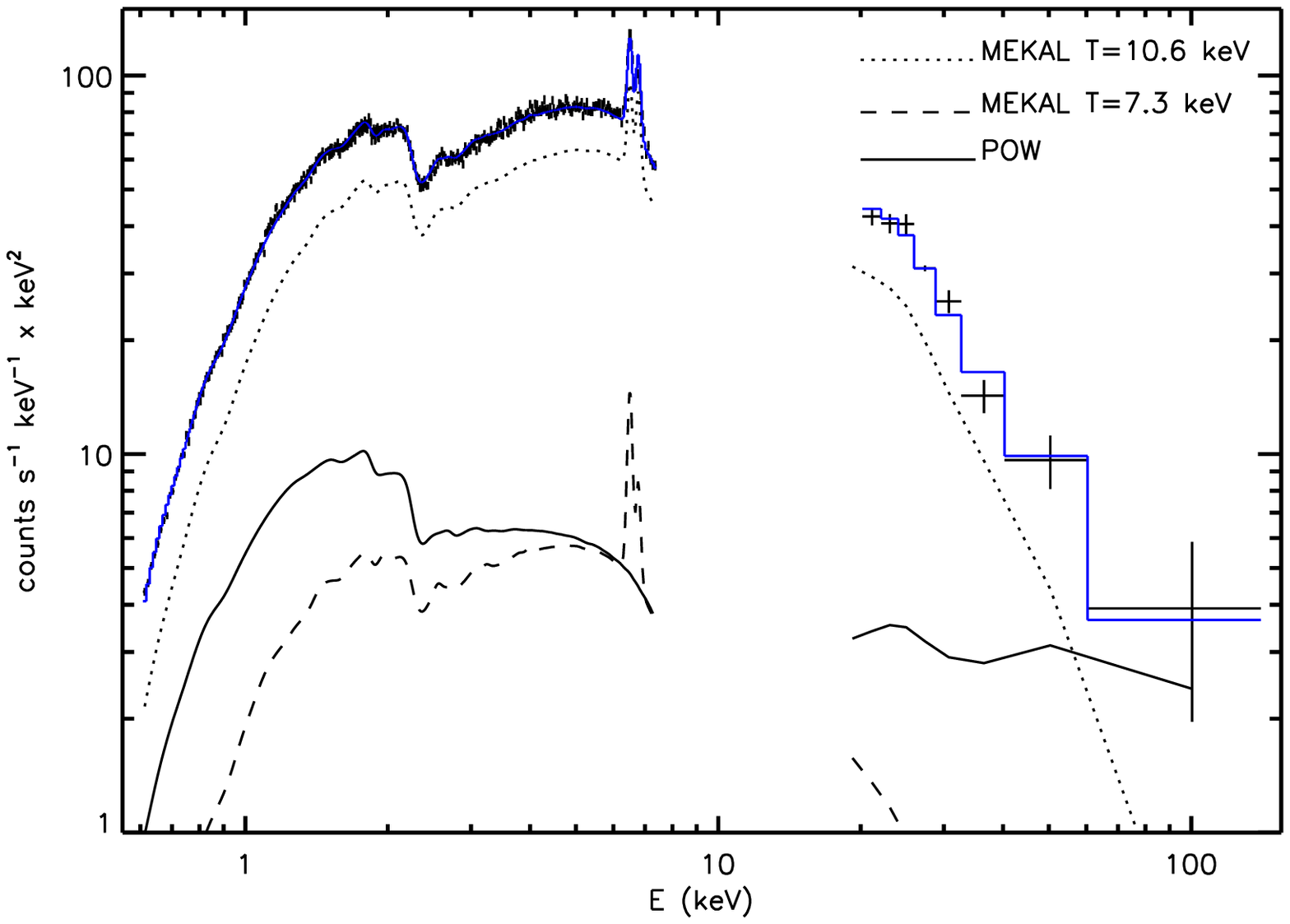}}
\caption{The black crosses show the data and the statistical uncertainties. 
In the left panel only the ISGRI data above 40 keV energies are shown, together with the prediction of the 
3T mekal + power-law model with different values for the photon index. 
In the right panel the PN and ISGRI spectra have been multiplied by E$^{2}$ for display purposes.
The blue line shows the total best-fit convolved and absorbed model with $\alpha_{ph} \equiv 2.3$. 
The following components of 
the total model are also shown separately: power-law (solid line), hot thermal component (dotted 
line) and the thermal component due to the cooling in the centre (dashed line).
\label{pow23.fig}}
\end{figure*}

We thus proceeded by fitting the data while fixing the photon index to a value in the range [1.5--3.0], 
allowing NH and the hot emission component parameters to vary.
We visually examined how well each of the forced best-fit models matches the ISGRI data.  
We found that models with $\alpha_{ph}$ = 1.5 are too hard compared to the ISGRI data.
With higher photon index the models fit the ISGRI data better, until they underpredict the data. The 
best agreement with the ISGRI data is obtained in the range 2.2--2.5 (see Fig. \ref{pow23.fig}). 
These models fit the PN data significantly better than the 3T models fitted to PN data only (see Section \ref{3tmodel}):
the overall residuals are smaller and the $\chi^{2}$/dof decreases significantly from 800.95 to 690--693 for 1 additional free parameter and 500 data bins.

\subsubsection{Results}
\label{results}
We will use the above models in the following to present the results. The presented ranges of the parameters are not
statistical 1 $\sigma$ uncertainties, but rather show the variation of the best-fit parameters in the models with 
$\alpha_{ph}$ fixed to range 2.2--2.5. 
Assuming that the IC and radio emission are produced by the same population of relativistic electrons 
(e.g. Sarazin \& Lieu, 1998) i.e.,
\begin{equation}
\label{ngamma}
N(\gamma) = N_{1} \times \gamma^{-p},
\end{equation} 
the implied differential momentum spectra of the relativistic electrons has a slope p of 3.4-4.0.
The corresponding range of the radio spectral index is 1.2--1.5. This is consistent with 
the estimated upper limit of 1.7 for the radio spectral index in Ophiuchus (P\'erez-Torrez et al., 2009).
The implied radio index $\sim$1.5 is at the border of that predicted by the primary and secondary models 
(e.g Ferrari et al., 2008) and thus we cannot distinguish those models with the current data.

In these models the flux of the power-law component in the 1--10 keV band is $\sim$10\% of the total.
The 20--80 keV band flux is 0.5--1.5 $\times 10^{-11}$ erg s$^{-1}$ cm$^{-2}$ .
The BeppoSAX PDS analysis of Ophiuchus (Nevalainen et al., 2004), while covering a much larger region,  
yielded a consistent 1$\sigma$ upper limit of 1.6 $\times 10^{-11}$ erg s$^{-1}$ cm$^{-2}$ for the non-thermal flux
in the 20--80 keV band for $\alpha_{ph}$ = 2.0 (Nevalainen et al., 2004). 
The 90\% upper confidence limit obtained with BAT using $\alpha_{ph} = 2.0$ (converted to the 20--80 keV band) is 
0.9 $\times 10^{-11}$ erg s$^{-1}$ cm$^{-2}$ (Ajello et al., 2009). Our hardest model ($\alpha_{ph} \equiv 2.2$)
exceeds this value significantly, while our other accepted models agree with the BAT value.

Using the formula for the power emitted by an electron through inverse compton scattering (e.g. Rybicki \& Lightman, 
1979) and using the electron distribution of Eq. \ref{ngamma}, one obtains a relation between the IC luminosity
and relativistic electron distribution as 
\begin{equation}
\label{nnorm}
\frac{L_{IC}}{V} = \frac{4}{3} \ \sigma_{T} \ c \ u_{\gamma} \ N_{1} \frac{\gamma_{max}^{3-p} - \gamma_{min}^{3-p}}{3-p},
\end{equation}
where L$_{IC}$ is the luminosity of the IC component in energy range [$\gamma_{min}$, $\gamma_{max}$], 
V is the emitting volume,  
$\sigma_{T}$ is the Thompson cross section ($\approx$ 6.652 $\times$ $10^{-25}$ cm$^{2}$) and $u_{\gamma}$ is the energy 
density of  the CMB ($\approx 4.356 \times 10^{-13}$ erg cm$^{-3}$).
Our full energy band of 0.6--140 keV band corresponds to $\gamma_{min} \sim$ 800 and $\gamma_{max} \sim$ 13000.  
For Ophiuchus, the power-law component with a photon index of 2.3 yields a luminosity of L$_{IC} \sim 1.3 \times 10^{44}$ 
erg s$^{-1}$ in this band.
Approximating the emitting volume as a sphere of 7 arcmin radius, we use volume V = $1.5 \times 10^{72}$ cm$^3$.
Using p = 3.6, as given by $\alpha_{ph}$ = 2.3, Eq. \ref{nnorm} yields a normalisation N$_{1} \sim 2.7$ cm$^{-3}$ for Ophiuchus. 
Now the electron population (Eq. \ref{ngamma}) is fully solved and we can compute the 
pressure of the relativistic electron population (P$_{IC}$) using
\begin{eqnarray}
P_{IC} = \frac{E}{3V} = \frac{1}{3} \int_{\gamma_{min}}^{\gamma_{max}} \ N(\gamma) \ \gamma \ m_{e} \ c^{2} \ d\gamma = \nonumber \\
\frac{m_{e} \ c^{2}\ N_{1}}{3}  \int_{\gamma_{min}}^{\gamma_{max}}  \gamma^{1-p}\ d\gamma = 
\frac{m_{e} \ c^{2}\ N_{1}}{3} \frac{\gamma_{max}^{2-p} - \gamma_{min}^{2-p}}{2-p}
\end{eqnarray}
yielding a value 1.1 $\times$ $10^{-12}$ erg cm$^{-3}$. On the other hand, the pressure of the thermal electrons is
given
by P$_{ther}$ = n$_{e,ther}$ kT$_{gas}$. The values within r=7 arcmin sphere in Ophiuchus 
(n$_{e,ther}$ $\sim 6.4 \times 10^{-3}$  cm$^{-3}$ and T$_{gas}$ $\sim$ 9 keV) yield P$_{ther} \sim 9.2 \times 10^{-11}$ 
erg cm$^{-3}$. Thus the pressure of the non-thermal electrons is $\sim$ 1\% of that of the thermal electrons.
Thus there is no problem of confining the relativistic electron population in the gravitational potential of the cluster.

\begin{figure*}[t]
\hbox{
\includegraphics[width=8cm,angle=0]{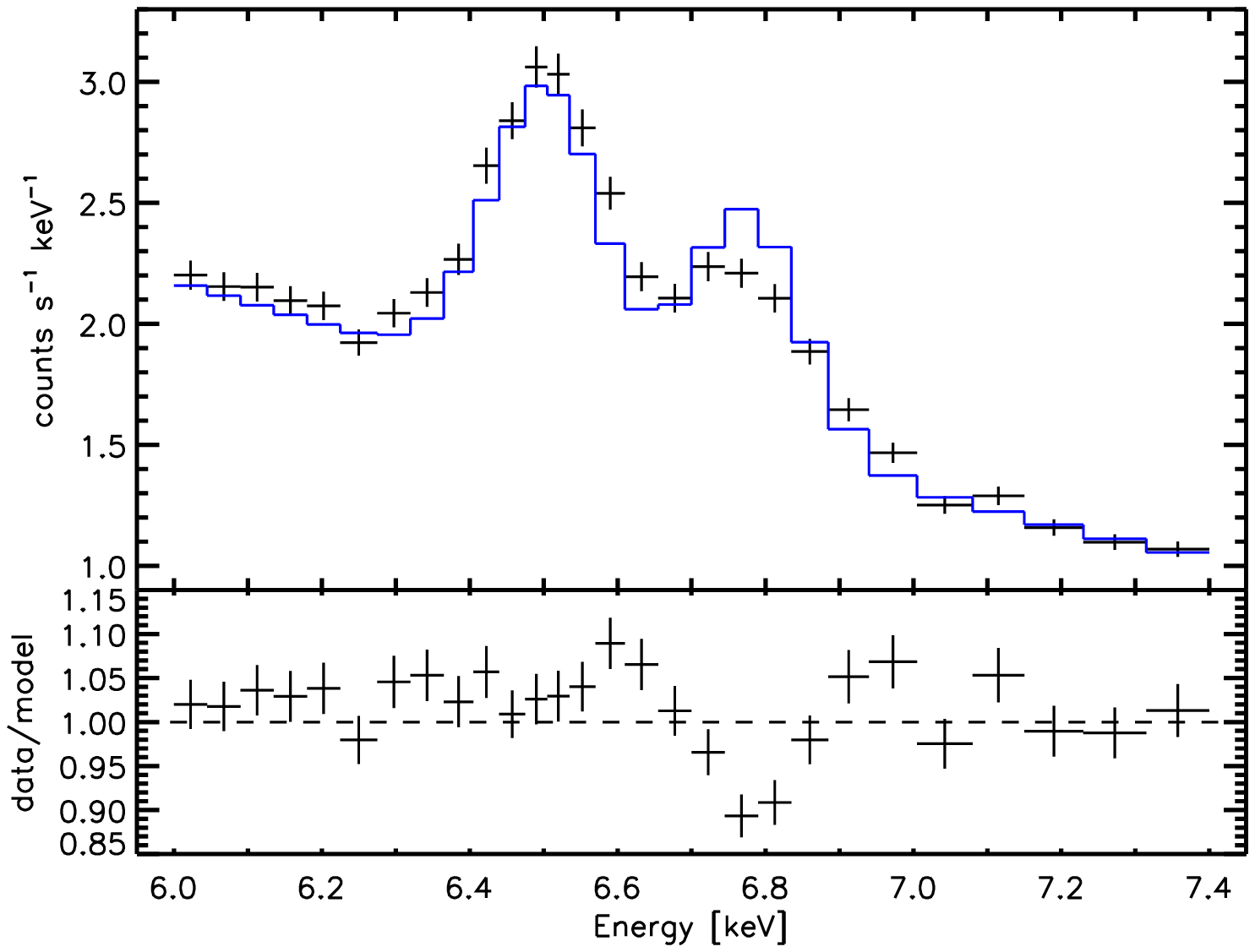}
\includegraphics[width=8cm,angle=0]{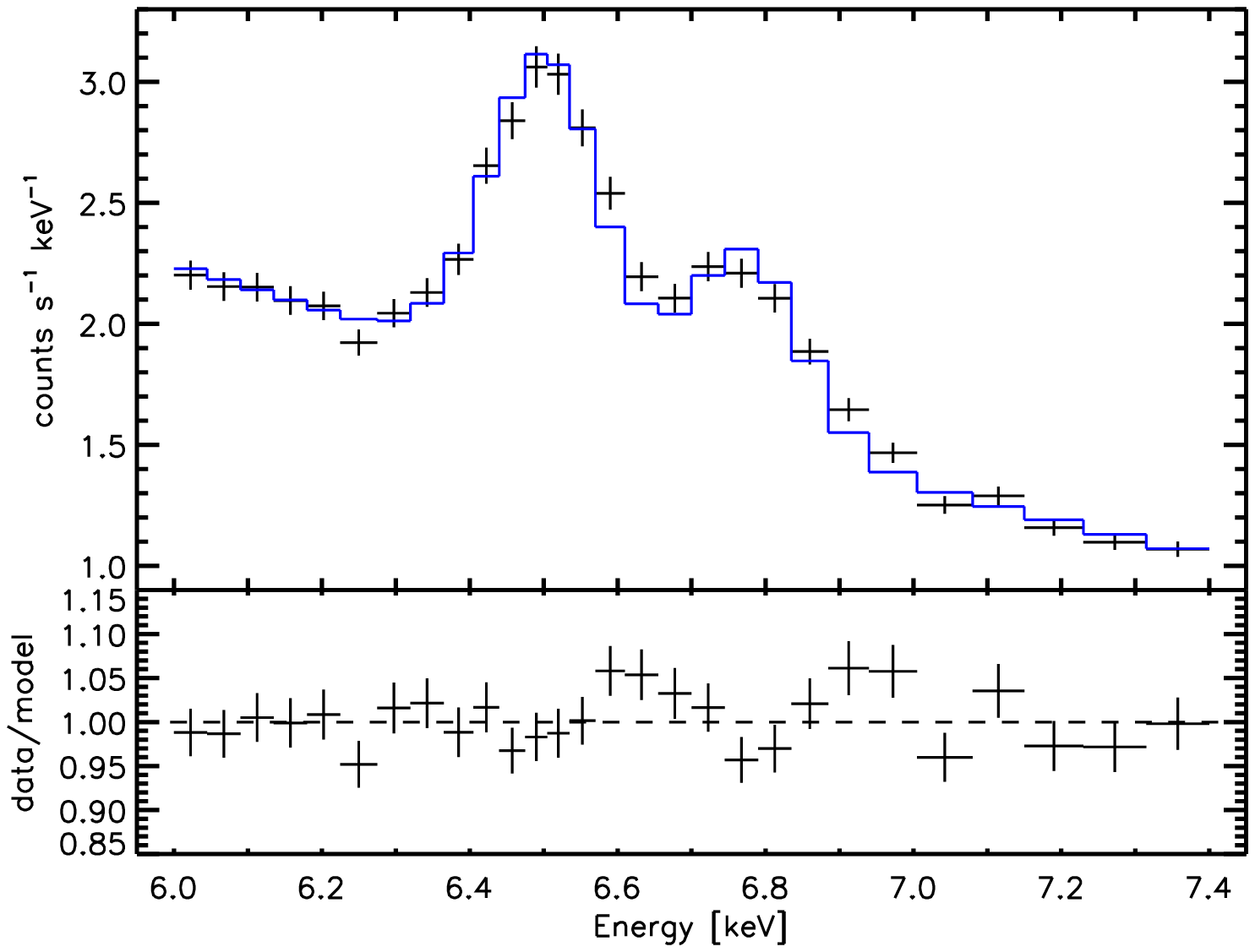}
}
\caption{The crosses in the upper panels show the PN spectrum of the central r=7 arcmin region in the 6.0--7.4 keV band. 
The blue line shows the convolved best-fit model when fitting the 0.5--100 keV band with a 3T+pow model with 
$\alpha_{ph} \equiv  2.3$ (left panel) and when fitting the 6.0--7.4 keV band with a single-temperature model (right panel)
. The lower panels show the ratio of the data and the model prediction 
\label{fe.fig}}
\end{figure*}

The VLA observations of Ophiuchus at 1.4 GHz yielded the flux of the mini-halo (see Fig. \ref{mos2image.fig}) 
as $\sim$ 106 mJy within a radius 9.6 arcmin (Govoni et al., 2009; Murgia et al., 2009). 
Using the measured exponential profile for the radio surface brightness in the above work, we obtained a mini-halo 
brightness of 85 mJy within our extraction region of r=7 arcmin.
We used this value together with the properties of the above power-law models to calculate the magnetic field 
strength using the formulation of Sarazin (1988).
The calculations yield B=0.05--0.15 $\mu$G, consistent with those derived in several other clusters (e.g. Rephaeli et al., 2008).

Due to the lack of a major merger in mini-halo clusters, two possibilities remain for the acceleration mechanism of 
electrons (e.g. Ferrari et al. 2008): i) primary old merger and subsequent re-accelerations due to turbulence or 
ii) secondary hardonic collisions. 
In Ophiuchus, the secondary acceleration models due to hadron-hadron collisions are inconsistent with the available 
hard X-ray measurements (Colafrancesco \& Marchegiani, 2009). 
A viable scenario for the Ophiuchus cluster is that the primary merger happened sufficiently long time ago (as required
by the long cooling time) so that most merger signatures have disappeared and that the relativistic electron population 
was later re-accelerated by the MHD turbulence. This is further supported by the steepness of the implied radio index
(1.2--1.5) which suggests that the electron population has experienced significant ageing. 
The relatively low magnetic strength derived in this work is qualitatively consistent with the primary/turbulence model: 
in a strong shock the magnetic field is amplified, so we would expect higher B field values. With time, the magnetic 
field decays and the particles move away from the shock regions, so one would expect lower magnetic field values in case 
of an aged population.

\section{Discussion}
\label{disc}
When we added the power-law component to the model (see Section \ref{results}), the 
resulting temperatures of the hot component 
(10.5--10.9 keV) became much higher than those derived with the Fe XXV/XXVI ratio ($\sim$ 9 keV, see 
Section \ref{spectralfitting}). We verified with simulations that the presence of the power-law component with the above parameters 
does not bias the temperature measured with the Fe lines.
The Fe XXV line emission is underpredicted while the Fe XXVI line emission is overpredicted, both 
at 10--15\% level, with the models presented here (see Fig. \ref{fe.fig}).
We note that when we fitted only the 6.0--7.4 keV band with a single-temperature model, some residuals still remained
at a 5\% level  (see Fig. \ref{fe.fig}). 
This indicates some uncertainties in the modelling of the energy redistribution or reconstruction of the 
photon energy (addressed elsewhere), or that the Fe line ratio deviates from the thermal prediction.

We experimented with the XSPEC model gsmooth, as Molendi \& Gastaldello (2009) in case of Perseus, to see 
whether the suggested mis-calibration of PN energy resolution is affecting the fits. Using the proposed values (4 eV 
at 6 keV for the Gaussian width and -1 for the power-law index) we obtained a very small change in the residuals, and 
in the direction of increasing the $\chi^{2}$. Allowing the Gaussian width to vary does not improve the fit. Thus, the 
possible miscalibration of the energy resolution does not produce the residuals.
This indicates at least two possibilities: either the real Fe XXV/XXVI line ratio is bigger than the thermal prediction,
or the assumed IC/CMB scenario or its modelling is wrong.

Our adopted method of simple addition of a power-law component to the emission model ignores the possible effect of 
the non-thermal electrons to the ionisation balance.
For instance, Prokhorov et al. (2009) showed that a presence of a population of supra-thermal electrons would decrease
the Fe XXV/XXVI flux ratio compared to that in the case of pure Maxwellian velocity distribution of electrons 
(as assumed in our work). This model may also produce a continuum that is different from the power-law. 
Note that this mechanism does not produce the (observed) synchrotron emission and thus we still need the IC/CMB 
mechanism at a lower level. A detailed analysis of this possibility (i.e. the co-existence of thermal, supra-thermal and 
relativistic electrons in the cluster volume) will be carried out in another paper.

Alternately, assuming that the Fe XXV/XXVI line ratio is correctly reproduced by the MEKAL model,  
we need to modify the modelling of the non-thermal component. 
The power-law shape itself should be the correct model in the IC/CMB case, since
the assumed diffusive shock acceleration process (Bell, 1978) does produce 
relativistic electron populations with a power-law distribution.
Also the radio spectra in cluster haloes do have power-law shape (e.g. Ferrari et al., 2008). 
Thus, we need to search for different physical mechanisms than IC/CMB that would produce the non-thermal emission 
spectrum different from a power-law.

\section{Conclusions}
We analysed the central r=7 arcmin region of the Ophiuchus cluster of galaxies using data obtained with the XMM-Newton 
EPIC and INTEGRAL ISGRI instruments.
The ISGRI data yielded a 5.7$\sigma$ detection of excess emission in the 20--120 keV band over the thermal prediction, as
modeled with PN data. 
Our XMM-Newton analysis confirmed the existence of a cool core in Ophiuchus. 
The derived very long cooling time (3 $\times 10^{9}$ yr) as well as the lack of significant merger signatures
argues against a recent major merger in the Ophiuchus centre.
These features are consistent with most clusters hosting a radio mini-halo (Ferrari et al., 2008), which has also been 
detected in Ophiuchus (Govoni et al. 2009). 

In most proposed models for the non-thermal emission in clusters the same population of relativistic electrons produces 
both the radio emission (via synchrotron) and hard X-ray emission (via inverse compton scattering of the cosmic microwave 
background photons). Our results support this scenario, since the derived photon index of 2.2--2.5 for the 
non-thermal component corresponds to radio index of 1.2--1.5 with is consistent with the upper limit (1.7) derived from 
the radio observations of Ophiuchus (P\'eres-Torrez et al.,  2009).

The non-thermal component produces $\sim$10\% of the total flux in the 1--10 keV band.
These models imply a differential momentum spectrum of the relativistic electrons with a slope of 
3.4-4.0 and a magnetic field strength B=0.05--0.15 $\mu$G.
The pressure of the non-thermal electrons is $\sim$ 1\% of that of the thermal electrons, i.e. the gravitational potential
of the cluster is adequate for confining such a population of non-thermal electrons.

\begin{acknowledgements}
JN is supported by the Academy of Finland. We thank F. Govoni, M. Murgia and C. Ferrari for providing the radio image. 
We thank K. Mattila for his help on the NH issue. We thank J. Wilms and D. Prokhorov for helpful discussions.
\end{acknowledgements}

\end{document}